\documentclass[aps,pra,showpacs]{revtex4-2}
\usepackage[english]{babel}
\usepackage[utf8]{inputenc}
\usepackage{graphicx}
\usepackage{amssymb}
\usepackage{xcolor}
\usepackage{amsmath}
\usepackage[utf8]{inputenc}
\usepackage{hyperref}
\usepackage{url}
\usepackage{color}
\usepackage{float}
\usepackage{enumerate}

\begin{document}

\title{Engineering Perfect State Transfer Graphs via Givens Transformations}

\author{Pablo Serra$^{(1)}$$^{(2)}$}
\author{Alejandro Ferrón$^{(3)}$$^{(4)}$}
\author{Omar Osenda$^{(1)}$$^{(2)}$}

\affiliation{${(1)}$ Facultad de Matemática, Astronomía, Física y Computación, Universidad Nacional de Córdoba, Av. Medina Allende s/n, Ciudad Universitaria, CP:X5000HUA Córdoba, Argentina,  
${(2)}$ Instituto de Física Enrique Gaviola (CONICET-UNC), ${(3)}$  Facultad de Ciencias Exactas y Naturales y Agrimensura, Universidad Nacional del Nordeste, Avenida Libertad 5470 , Campus Deodoro Roca, W3402 Corrientes, Argentina, and ${(4)}$ Instituto de Modelado e Innovación Tecnológica, (CONICET-UNNE)}
 
\begin{abstract}
Perfect quantum state transfer is achievable in different settings, including linear qubit chains, bi-dimensional arrays, ladders, etc. The most studied case contemplates transferring arbitrary one-qubit pure states in systems with homogeneous interactions. These restrictions allow finding numerous examples of systems that show perfect transfer but in geometries that are not implementable or are very difficult to implement in actual experimental settings. Relaxing the homogeneity of the interactions and inspired by the $XX$ qubit chains that show perfect transmission, we present a simple scheme based on the Givens Transformations to analyse and obtain a class of qubit graphs that possess perfect quantum state transmission. We present some simple examples and show how it is possible to generalize them for longer transmission lengths. 
\end{abstract}

\maketitle

\section{Introduction}\label{sec-intrro}

Perfect quantum state transfer \cite{Xie2023,Maleki2021,Mograby2021,Christandl2004,Li2018, Bayat2014} is one of the simpler tasks studied in the exceedingly ample field of quantum information processing. The transfer of arbitrary quantum states and quantum computation have the same prescriptions to proceed. First, prepare an initial state. Second, let the state evolve by following the autonomous or controlled dynamic evolution given by the system Hamiltonian and its environment. At last, measure the final state to check the result. The difference between transfer and computation concerns the expected product of the procedure. In the case of quantum state transfer, the result should be the recovery of a transmitted state with a strong resemblance to the initial state, not the unknown result of some algorithm. 

Since the first studies on the area \cite{Bose2003,Bose2008}, quantum state transfer has been considered an operation involving the communication of two processors or two disconnected parts of a processor. So, most studies focused on linear geometries or chains of qubits \cite{Serra2022JPA,Wang2022,Yousefjani2021a,Zwick2015}. This trend has changed, in part thanks to the development of quantum processors with different connection patterns, as is the case with the processors based on superconductor qubits \cite{Li2018,Serra2025}. So, studying PT in graphs becomes a necessity.  

Besides the geometrical patterns in the connectivity between qubits in quantum processors, there are other reasons to study QST in qubits graphs \cite{Mograby2021,Godsil_2011,Fan2013,Kempton2017,Cheung2011,Ge2011,Coutinho2022}, for instance, the interest in the transmission of more complicated states than arbitrary one qubit states, such as entangled states shared between few or several qubits, more resilient communication channels, etc. Since many studies assume that the interactions between the qubits on the graphs are homogeneous, it is possible to resort to a trove of theorems about the spectral properties of the graph's Hamiltonian \cite{waterloo2022,waterloo2022b}. Then, if the spectrum of a given graph Hamiltonian fulfils the conditions necessary for the appearance of PT, the library of systems that offer PT is augmented. Nevertheless, very often, the spectrum of the Hamiltonian of given graphs of qubits satisfies only partially the necessary conditions, and, instead of PT, the onlooker finds pretty good transmission \cite{Fan2013,Serra2025} which allows to reach transmission fidelities as close to unity as required, but in a time scale incompatible with practical implementations \cite{Serra2024}. 

While the number of examples of graphs allowing PT or PGT \cite{Coutinho2015,Himmel2025,Himmel2025b} is slowly growing, the graphs obtained show practical issues. Sometimes the geometry cannot be implemented in actual processors, as in tree-like settings. The scenario described leads us to pursue a simple method that allows us to assess when a graph shows PT by removing the condition about the homogeneity of the interactions. Consequently, we want to study graphs in which nodes that have three or more links are not connected to each other, nor share any of those links among themselves, considering the latest designs used in quantum processors \cite{Bravyi2022,heavyhex, Kattemolle2025}. Another restriction in the present work is that we focus on graphs whose qubits interact through nearest-neighbour interactions, and the Hamiltonian of the system only contains ferromagnetic $XX$ type interactions. Finally, we concentrate on the transmission of one-qubit quantum states.

In general, we say that a graph with $N$ qubits is exactly solvable if we can get an analytical algebraic expression for the $N$ eigenvectors and eigenvalues of the restriction of the Hamiltonian to the one-excitation subspace. In this case, checking for perfect transmission becomes a simple task. 

The restriction of the Hamiltonian to the one-excitation subspace does not result in a tridiagonal matrix for general graphs, so the number of results or theorems applicable to ensure the existence of exact expressions for the spectrum is poor. This scenario becomes worse when the strength of the interactions is site-dependent. There are well-known cases of chain Hamiltonian matrices with an exact spectrum compatible with perfect quantum state transfer, and their Hamiltonian matrices are tridiagonal matrices with null diagonal entries \cite{Christandl2004,Karbach2005,acde2004}. So, it seems reasonable to transform the Hamiltonian matrix of the graph into the Hamiltonian matrix of a chain using an isospectral transformation \cite{VENKATESHAN201419}. Then, select the values of the chain's matrix entries to make them compatible with perfect transmission. This results in equations for the entries of the graph Hamiltonian matrix that depend on both the values of the entries of the chain Hamiltonian matrix and the transformation coefficients. Of course, several transformations lead any real symmetrical matrix into a tridiagonal one, and almost any numerical diagonalization algorithm to calculate the spectrum employs this kind of transformation \cite{lapack,numericalrecipes}. 

To obtain a proper recipe applicable to distinct graphs it is necessary to contemplate some inconveniences, starting with which are the required conditions that ensure obtaining null diagonal entries on the chain Hamiltonian matrix, that the equations for the graph Hamiltonian matrix entries have algebraically exact solutions and both, the similarity transformation and its inverse can be analytically and exactly implemented. 

Although meeting all the requirements above seems a tall order,  we demonstrate that the procedure works for several particular examples and that several necessary conditions to achieve it arise naturally.

We organised the paper as follows. In Section~\ref{sec-graphs-givens}, we present the Hamiltonian that describes the interactions between the qubits on an array and the relationship between an arrangement of qubits and the underlying graph that represents it by locating the qubits on the nodes of a graph and depicting and the interactions between the qubits as the links between the nodes of the graph. We also present the Givens Transformation \cite{numericalrecipes}, which is the principal tool that we use to investigate the properties of different graphs. It is worth mentioning that the Given Transformations (GT), also known as Givens Rotations, are ubiquitous in diverse contexts that involve matrix calculations or applications, ranging from solving linear equations \cite{lapack} to fine-tuning of large language models \cite{Ma2024}. GTs play a significant role in Quantum Computations, allowing the decomposition of complicated operations into elementary unitary operations \cite{Cybenko2001}.

In Section~\ref{sgcs}, we employ the GT to analyse a graph introduced some years ago by Gabriel Coutinho \cite{coutinho2016}. The qubit system related to this graph presents perfect transmission for appropriate interaction coefficients. We demonstrate how these particular values of the interaction coefficients reappear once we transform the one-excitation Hamiltonian matrix of the graph with variable coefficients into the $XX$ Hamiltonian matrix of a linear chain. We also present two other graphs, closely related to Coutinho's, to further clarify our procedure. Then, in Section~\ref{sec-gen-lemmas}, we present three other graphs with particular architectures that show perfect transmission for an arbitrary number of qubits. We formulate and demonstrate three Lemmas, one for each architecture, which provide the interaction coefficients compatible with perfect quantum state transmission.  

In Section~\ref{sec-bridges} we explore the results of adding extra links, or bridges, to a graph that already has perfect quantum state transmission. By doing so, we find that the transformation of the Hamiltonian matrix of a graph with an added bridge to the Hamiltonian matrix of a linear chain could hinder the appearance of PT, or that, in some cases, to recover the PT, it is necessary to admit positive and negative interaction coefficients. After considering all the examples, we arrive at the necessary conditions that warrant perfect state transmission. We will ask that the graph under consideration be bipartite \cite{waterloo2022b}.  We also analyse a decorated chain implementable using transmon qubits, which is exactly solvable \cite{Serra2025, Serra2024}.

We summarise our findings and draw some conclusions in Section~\ref{sec-conclusions}. For completeness, we present some lengthy algebraic manipulations in the Appendices. 

\section{Spin Chains, Graphs and Givens Transformation}\label{sec-graphs-givens}

One of the most valuable constructions of physics is the concept of lattice, {\em i.e.}, an ordered array of sites having a basis set of vectors endowed with symmetry properties such as translational symmetry, reflection symmetry and so on. In most situations, the lattice's mathematical properties help to calculate physical quantities. Graphs, on the other hand, are similar structures but allow the introduction of a variable number of neighbours, do not imply the notion of distance between the sites of the graph, etc. Graphs are a collection of sites and links between them. Both elements enter into the adjacency matrix. The connection between graphs and qubits arrays is simple. The qubits lie at the sites of the graph, and the links show what qubits interact with other qubits.

\begin{figure}[hbt]
\includegraphics[width=0.7\linewidth]{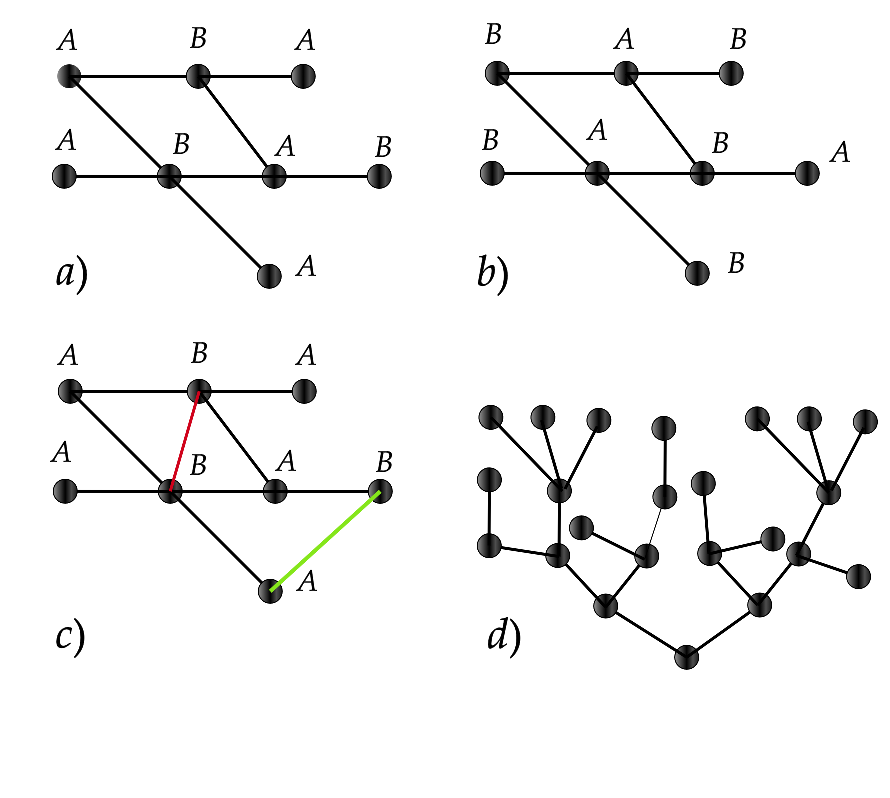}
\caption{The cartoon depicts some properties of graphs. The circles and the straight lines correspond to the nodes and links of the graph, respectively. a) and b) show the same bipartite graph with two different labellings. The bipartite condition does not depend on the particular label employed on a single node. Labelling any single node as “A” or “B” and then labelling all nodes connected to it as “B” or “A,” respectively, makes it possible to assign labels to all nodes in the graph uniquely. c) shows that the addition of links to a bipartite graph leads to the loss of that condition if the link added is the red one. If the link added is the green one, the graph remains bipartite. Note that the nodes connected by the red link have three nodes connected to them before the introduction of the red bridge. d) shows a tree-like graph. The nodes on the graph belong to a layer, and the number of links that separate a given node from the original one labels the corresponding layer. The distance between two nodes is the smallest number of links needed to join them.  Note that tree-like graphs are always bipartite. }\label{fig-cartoon}
\end{figure}

The cartoon in Figure \ref{fig-cartoon} shows several graphs where the circles correspond to the sites and the straight lines to the links joining them. Figure \ref{fig-cartoon} a) and b) show the same bipartite graph and the two ways of labelling the sites. In each case, an $A$ ($B$) site is connected only to $B$ ($A$) sites that are their closest neighbours, not in a physical distance but in the number of links that separate the site from its nearest neighbours. In a graph,  the number of links of the shortest path joining two sites replaces the physical distance in a lattice. Figure \ref{fig-cartoon} c) shows how adding links could or could not break the bipartite condition. Adding the green link does not change the bipartite condition since it adjoins two sites with different labels while adding the red link destroys the bipartite property. Figure \ref{fig-cartoon} d) shows a tree-like graph with branching equal to two. 

The number of sites, the number of links departing from each site, and the pairs $(i,j)$ denoting which pairs of sites are connected determine univocally a graph. Arranging the list of connected pairs results in the well-known adjacency matrix. The more general $XX$ Hamiltonian for a graph is

\begin{equation} \label{ec:ssh-hamiltonian}
H= - \sum_{(i,j)\in E} \left( J_{ij} \sigma_{i}^+ \sigma_j^- + h.c. \right)+\sum_{i\in {\cal V}} B_i\sigma_i^z \, ,
\end{equation}

\noindent where $\sigma$ are the Pauli operators, the $J_{i,j}=J_{j,i}$ are the interaction strengths, $B_i$ is a local field applied to the $i$ site, $E$ is the set of links, ${\cal V}$ is the set of vertices, and the first sum runs over the pairs $(i,j)$. 

The Hamiltonian for the one excitation subspace results in a real symmetrical matrix. For instance, for the graph depicted in Figure \ref{fig-cartoon} a), the one-excitation Hamiltonian matrix reads,

\begin{equation}\label{eq-hamiltonian-matrix-8}
h =
\left(
\begin{array}{cccccccc}
B_1     & J_{1,2} & 0 & 0 & J_{1,5} & 0 & 0 & 0  \\
J_{1,2} & B_2    & J_{2,3} & 0   & 0  & J_{2,6} & 0 & 0 \\
0  & J_{2,3} & B_3 & 0& 0 & 0 &  0 & 0  \\
0  &  0 & 0  & B_4  & J_{4,5} & 0 & 0 & 0 \\
J_{1,5} & 0 & 0 & J_{4,5} & B_5 & J_{5,6} & 0 & J_{5,8} \\
0 & J_{2,6} & 0 & 0& J_{5,6} & B_6 & J_{6,7} & 0  \\
0 & 0 & 0 & 0& 0 & J_{6,7} & B_7 & 0  \\
0 & 0 & 0 & 0 & J_{5,8} & 0 & 0 & B_8
\end{array}
\right).
\end{equation}

\noindent  We number the sites of the graph in Figure \ref{fig-cartoon} a) using that the sites belong to one of three rows. The first row has three sites, the second four, and the third just one site, numbering the sites from left to right. Following this, the first site labelled with a $B$ on the second row is the fifth site of the graph and connects with the first, fourth, sixth and eighth sites. This also can be appreciated by looking at the fifth row of the matrix, Equation \ref{eq-hamiltonian-matrix-8}. The matrix in Equation \ref{eq-hamiltonian-matrix-8} is not tridiagonal, which hampers the chance of finding exact analytical and algebraic expressions for the eigenvalues and eigenvectors unless we transform it into a tridiagonal one. In the single-excitation subspace, a tridiagonal matrix is always equivalent to the Hamiltonian of a linear chain with open boundary conditions. Moreover, for such linear chains, the conditions required for the emergence of perfect transmission (PT) or pretty good transmission (PGT) are well known \cite{Christandl2004,kay2010,Serra2024}. For instance, a linear chain with coupling coefficients given by 
\begin{equation}
\label{e-cpt}
J_n=\sqrt{n\,(N-n)}, 
\end{equation}
where $N$ is the chain length, 
exhibits perfect transmission, although other configurations can also guarantee PT.

Note that external field terms, depending on the $B_i$ coefficients, appear only in the diagonal of the Hamiltonian matrix, so, in general, we call any diagonal entry of a Hamiltonian matrix a "field term". On the other hand, any graph with the property that any site on it connects only with one or two different sites of the graph is a chain. If it is allowed to introduce sites with three links, but no sites with three links connect directly, and there are no closed paths on the graph, a decorated chain is obtained, as is the case of decorated SSH chains, also known as generalized SSH chains.  

As said in the previous Section, the usual way to numerically calculate the eigenvalues of real symmetrical matrices consists of transforming them into symmetrical tridiagonal matrices. Numerical packages like Lapack \cite{lapack} or even the Numerical Recipes \cite{numericalrecipes} use the Householder transformation because of its numerical efficiency. Householder transformation has a severe drawback for our purpose. The identity is not a Householder transformation, so even tridiagonal matrices change under this type of transformation. Instead, the Givens transformation (GT) leads any real symmetric matrix to a tridiagonal shape and reduces to the identity when the target matrix is already tridiagonal. It is similar to a rotation in a two-dimensional subspace. 

A Givens rotation matrix $G(i,j,\theta)$ is an $n \times n$ orthogonal matrix that acts on a given pair of coordinates $(i,j)$ in an $n$-dimensional vector while leaving all other coordinates unchanged and it has the form:

\[
G(i,j,\theta) = I_n + (c-1) e_i e_i^T + (c-1) e_j e_j^T + s e_i e_j^T - s e_j e_i^T
\]

\noindent where $I_n$ is the $n \times n$ identity matrix, $e_i$ and $e_j$ are the standard basis vectors, $c = \cos\theta$, and $s = \sin\theta$. Explicitly, $G(i,j,\theta)$ is an identity matrix with some modifications:

\[
G_{ii} = c, \quad G_{jj} = c, \quad G_{ij} = s, \quad G_{ji} = -s.
\]
When the GT is applied to a vector $x \in \mathbb{R}^n$, the transformation rotates the $(i,j)$-plane coordinates keeping the other components unchanged:

\[
G(i,j,\theta) x =
\begin{bmatrix}
    x_1 \\
    \vdots \\
    c x_i + s x_j \\
    \vdots \\
    -s x_i + c x_j \\
    \vdots \\
    x_n
\end{bmatrix}
\]

GTs introduce zeros in a matrix when performing QR decomposition. Given a matrix $A \in \mathbb{R}^{m \times n}$, we can iteratively apply Givens rotations to transform it into an upper triangular form $R$, such that:
\[
Q^T A = R,
\]
\noindent where $Q$ is the product of Givens rotations and is an orthogonal matrix. This is simpler when the Hamiltonian matrix is real and symmetric, in which case Givens transformations always reduce it to a tridiagonal form.

\section{Coutinho Graph as a starting point \label{sgcs}}

\begin{figure}[hbt]
\includegraphics[width=0.4\linewidth]{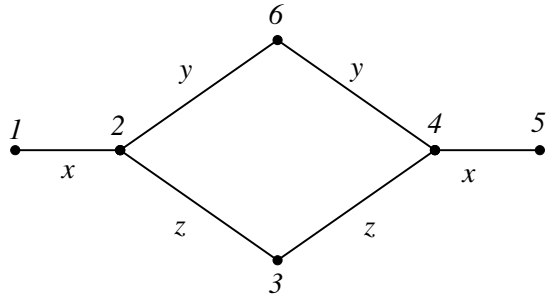}
\caption{The cartoon depicts the graph proposed in Reference~\cite{coutinho2016}. The numbering of the nodes helps identify the relationship between the interaction coefficients and the entries of the Hamiltonian matrix. See Equation \ref{eq-coutinho-6}. The distance between nodes 1 and 5 is equal to 4, and the coefficients are
$x,\,y,\,z$.\label{fcou}}
\end{figure}

The study of graphs in this context encompasses a huge set of spin arrays. Among them, the simplest and most fundamental structure is the linear chain,  a cornerstone of this work. From this basic configuration, one can introduce various modifications to position the spins and their interactions, leading to increasingly complex graph structures. However, our aim is not to start with the most intricate or convoluted graphs but, instead,  to build upon simple, well-understood foundations. Coutinho \cite{coutinho2016} identified a small graph exhibiting perfect state transfer over a distance $D=4$. This graph was later analyzed by Kay \cite{kay2018}, who described it in detail. In Fig. \ref{fcou}, we show the simplest possible graph, which we will call the Coutinho graph. This graph, for $x=z=2$ and $y=\sqrt{2}$, has a perfect transmission with 

\[
P(t)\,=\,\sin^8(t) \,,
\]

\noindent and will be used throughout this section to explore more complex graphs.

The algorithm proposed by Kay \cite{kay2018} starts with a linear chain that exhibits perfect transfer (PT) and applies a node-splitting procedure that preserves the PT condition. This process allows us to generate more complex graphs, with the simplest case being the Coutinho graph shown in Fig. \ref{fcou}. We need to remark that this algorithm requires that the initial linear chain have PT and then determines the interactions that ensure PT in the resulting graph. In contrast, our approach wants to start with a graph with arbitrary interactions and reduce it to a linear chain. We then impose the PT condition on this linear chain, allowing us to determine all possible interaction values that map to a specific linear chain and not just a particular set. 

Using the numbering of Figure~\ref{fcou} for the qubits, and its labels for the interactions, we get the Hamiltonian matrix
\begin{equation}\label{eq-coutinho-6}
h_6 = 
\left(
\begin{array}{cccccc}
0  &  x  &  0  & 0 & 0 & 0 \\
x &  0 & z  & 0 & 0 & y \\
0 & z & 0 & z & 0 & 0 \\
0 & 0 & z & 0 & x & y \\
0 & 0 & 0 & x & 0 & 0 \\
0 & y & 0 & y & 0 & 0
\end{array}
\right)
\end{equation}

Note that the Coutinho graph, as presented in fig. \ref{fcou}, does not have mirror symmetry about the axis of the graph (although it does around an axis that passes through sites 3 and 6.). Nevertheless, due to
lemma 2 of ref. \cite{kay2010}, it is a necessary condition for having PT between the ends of an XY linear chain that it has mirror symmetry, i.e., $J_n=J_{N-n}$ and $B_n=B_{N-n+1}$ (in our case $B_n=0$). For this reason, we need to reduce the graphs with PT discussed in \cite{kay2010} to centrosymmetric XX chains. Applying GTs, we see that the graph \ref{fcou} reduces, after 5 GTs, to a 5-site centrosymmetric chain with

\begin{equation}
\label{ejc}
J_1\,=\,J_4\,=\,x \quad;\quad J_2\,=\,J_3\,=\,\sqrt{y^2+z^2} \,.
\end{equation}

\noindent and imposing the condition 

\begin{equation}
\label{eclcspt}
J_n=\sqrt{n\,(N-n)}, 
\end{equation}

\noindent the chain presents PT $\forall \;N$ \cite{acde2004}. We can see that the Coutinho graph (see Fig. \ref{fcou}) presents PT for $x=2, \;y^2+z^2=6$, and the one presented by Coutinho in \cite{coutinho2016,kay2010} results a particular case of the one obtained using GTs. As we noticed, Coutinho's graph is not symmetric, but we can modify it to be symmetric obtaining the same linear chain during the procedure. This condition for $x, y, z$ is not the only one that presents PT, there are infinite possibilities that also present PT even if they do not satisfy Eq. (\ref{eclcspt}). We can see that the following condition, much more general, also assures us PT for the Coutinho graph between the sites 1 and 5:

\begin{eqnarray}
\label{ecxyz}
x&=&2\,(2\,j+1), \nonumber \\
y^2+z^2&=&2\,(4 \,m^2-(2\,j+1)^2) \,, %x^2+2 (y^2+z^2)=4\,m^ 2 \;\Rightarrow\; 
\end{eqnarray}

\noindent where $m> j$ are integer numbers. 

\begin{figure}[hbt]
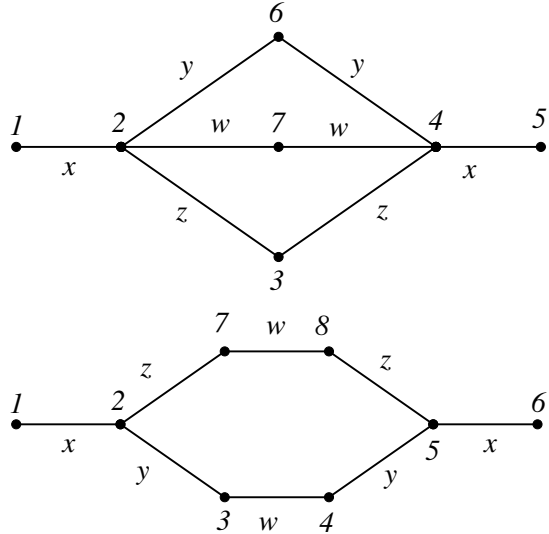

\includegraphics[width=0.4\linewidth]{tricoutinho.eps}\\
\vspace{0.25cm}
\includegraphics[width=0.4\linewidth]{coutinhoN8.eps}
\caption{ The cartoon in the figure shows two graphs with more nodes and links than the one originally proposed by Coutinho \cite{coutinho2016}. In panel (a), the largest distance between nodes is 4, while in panel (b) it is 5. \label{ftric}}
\end{figure}

Now, we can introduce modifications to the Coutinho graph to increase its complexity. We can perform this by adding spins either in series or in parallel. Adding sites in parallel increases the number of branches connecting the first and last sites, while adding sites in series preserves the original number of branches.
Figure \ref{ftric} illustrates these two approaches. In Fig. \ref{ftric} (a), we introduce an additional branch, increasing the total number of sites by one. In contrast, in Fig. \ref{ftric} (b) we observe a new graph that keeps the original number of branches from the Coutinho graph but adds two extra sites to the spin array. The first case represented in Fig. \ref{ftric} (a) preserves the distance $D=4$ of the original Coutinho graph while the one represented in Fig. \ref{ftric} (b) has distance $D=5$. 

The generalization of the Coutinho shown in Fig. \ref{ftric} (a),  the $N=7$-sites graph, is reduced to the same linear chain of $N=5$-sites as the original Coutinho graph, only by replacing in $J_2$ and $J_3 \quad y^2+z^2 \;\rightarrow w^2+y^2+z^2$. Doing this replacement in \eqref{ecxyz} we can obtain the PT condition for the new graph. Furthermore, we can observe that the graph of Fig. \ref{ftric} is first transformed into the original Coutinho graph \ref{fcou} and then into a linear chain, giving us an idea to explore induction demonstrations for more complex cases.

The case for $N=8$ and $D=5$ shown in Fig. \ref{ftric} (b) is more challenging. The one excitation Hamiltonian is written as,

\begin{equation}
\label{emh0cN8}
h = 
\left(
                 \begin{array}{cccccccc}
                  0 & x & 0 & 0 & 0 & 0 & 0 & 0 \\
                  x & 0 & y & 0 & 0 & 0 & z & 0 \\
                  0 & y & 0 & w & 0 & 0 & 0 & 0 \\
                  0 & 0 & w & 0 & y & 0 & 0 & 0 \\
                  0 & 0 & 0 & y & 0 & x & 0 & z \\
                  0 & 0 & 0 & 0 & x & 0 & 0 & 0 \\
                  0 & z & 0 & 0 & 0 & 0 & 0 & w \\
                  0 & 0 & 0 & 0 & z & 0 & w & 0 \\
                 \end{array}
                 \right) \,.
\end{equation}

\noindent and, after applying five GT we obtain the Hamiltonian corresponding to the 6-site centrosymmetric linear chain (see Appendix \ref{apn8} for more detail in the GT process):

\begin{equation}
\label{emhfcN8}
\left(
\begin{array}{cccccccc}
                  0 & x & 0 & 0 & 0 & 0 & 0 & 0 \\
                  x & 0 & \sqrt{y^2+z^2} & 0 & 0 & 0 & 0 & 0 \\
                  0 & \sqrt{y^2+z^2} & 0 & w & 0 & 0 & 0 & 0 \\
                  0 & 0 & w & 0 & \sqrt{y^2+z^2} & 0 & 0 & 0 \\
                  0 & 0 & 0 & \sqrt{y^2+z^2} & 0 & x & 0 & 0 \\
                  0 & 0 & 0 & 0 & x & 0 & 0 & 0 \\
                  0 & 0 & 0 & 0 & 0 & 0 & 0 & w \\
                  0 & 0 & 0 & 0 & 0 & 0 & w & 0 \\
                 \end{array}
                 \right)\,,
\end{equation}

\noindent  This $N=6$ linear chain presents PT when its interactions, given by Eq. \eqref{eclcspt}, are chosen such as $J_1=x=\sqrt{5}$, $J_2=\sqrt{y^2+z^2}=2\,\sqrt{2}$, and $J_3=w=3$. It is important to note that if, instead of defining a parameter $w$, we choose two different ones for $J_{34}$ and $J_{78}$, then the resulting linear chain is a chain with 8 sites. In this case, the expression of the parameters $J$ as a function of the parameters of the initial graph results very complicated. Furthermore, when we impose that the chain be centrosymmetric, all the solutions are with $J_{34}=J_{78}$.

After some analytical manipulation, we get that for the graphs in Figure~\eqref{ftric} b) the probability of transmission between the first and sixth qubits is

\begin{equation}
\label{ePcN8}
P_{16}\,=\,\sin^{10}(t)\,.
\end{equation}

\section{Generalization for the Coutinho Graph}\label{sec-gen-lemmas}

\begin{figure}[hbt]
\includegraphics[width=0.5\linewidth]{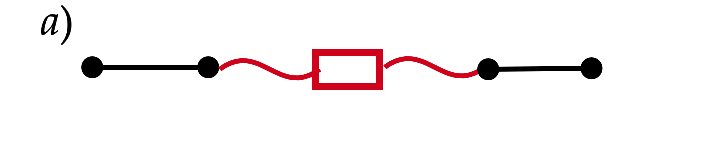}
\includegraphics[width=0.5\linewidth]{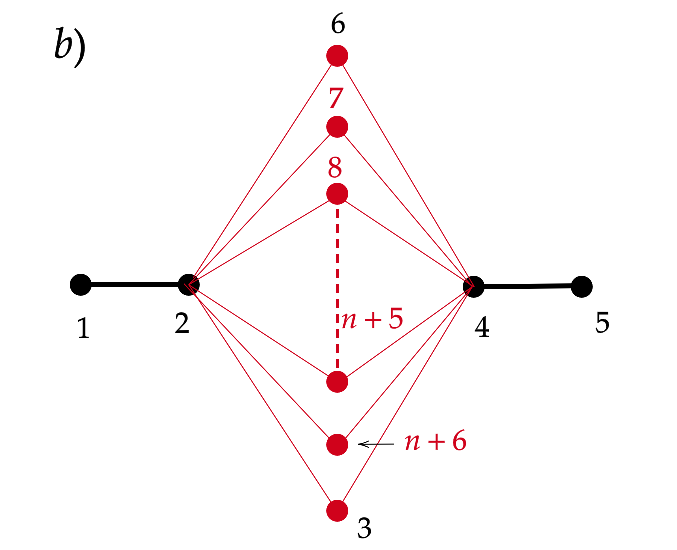}
\caption{\l(a) The cartoon shows how to generalize the graph originally proposed by Coutinho \cite{coutinho2016}. The two tails, each formed by two nodes connected by a link (depicted in black), are connected to an internal set of nodes and links, such that at most two links reach at each internal node. The internal nodes and links correspond to the red rectangle, and all links connecting the internal nodes to the innermost tail nodes are represented by red wavy lines. (b) An example in which $n+3$ links connect to the innermost nodes of the tails.\label{fig-generalizing-coutinho}}
\end{figure}

In this Section, we aim to present several generalizations of the Coutinho graph by changing what lies between the leftmost pair of qubits and the rightmost pair of qubits,  as the cartoon in Figure \ref{fig-generalizing-coutinho} the graphs in this Section have two "tails" formed by two pairs of qubits. In the upper panel, the two pairs connect with the rest of the graph through the red curvy links. The framed red box stands for the other sites and links of the graph, but the two pairs. The QST takes place between the outermost qubits of the tails, {\em i. e.} the state is prepared on the first qubit, which connects with only one qubit and the transfer is verified at one qubit that it is also connected only to another qubit of the graph, see Figure \ref{fig-generalizing-coutinho}. The lower panel shows a generalization that consists of joining qubits $2$ and $four$ with $n$ {\em bridges}, each bridge formed by two links and a single qubit shown using straight red lines and a solid red dot. \\

\noindent{\bf First generalization and Lemma}\\

\noindent {\bf Lemma 1.}  Let us consider the graph in Figure \ref{fig-generalizing-coutinho} b), which has $N=6+n$ qubits. Then, the one-excitation Hamiltonian matrix (OHM) of this graph reduces to the OHM of a chain with $5$ sites by applying a succession of Givens Transformations. It is possible to choose the matrix elements of the later Hamiltonian matrix to obtain PT in the five qubit chain and, therefore, on the graph with $N=6+n$ qubits. 

\noindent{\bf Proof.} By mathematical induction. It is enough to demonstrate that the OHM of a graph with $n+6$ qubits transforms to the Hamiltonian matrix of a similar graph but with $(n-1)+6$ qubits. We start up showing the reduction of the OHM for $n=1$, {\em i. e.} we show that the OHM of the graph depicted in Figure \ref{fcoutinhoN8} a) reduces to the OHM of the graph in Figure~\ref{fcou}. The OHM for the graph with $1+6$ qubits reads as

\begin{equation}
\label{eh7}
h_7=\left(
                 \begin{array}{ccccccc}
                  0 & x & 0 & 0 & 0 & 0 & 0 \\
                  x & 0 & z & 0 & 0 & y & w \\
                  0 & z & 0 & z & 0 & 0 & 0 \\
                  0 & 0 & z & 0 & x & y & w \\
                  0 & 0 & 0 & x & 0 & 0 & 0 \\
                  0 & y & 0 & y & 0 & 0 & 0 \\
                  0 & w & 0 & w & 0 & 0 & 0 \\
                 \end{array}
                 \right) \,.
\end{equation}

\noindent By applying the following GT 
\begin{equation}
\label{ep7}
P_7=
 \left(
   \begin{array}{ccccccc}
    1 & 0 & 0 & 0 & 0 & 0 & 0 \\
    0 & 1 & 0 & 0 & 0 & 0 & 0 \\
    0 & 0 & 1 & 0 & 0 & 0 & 0 \\
    0 & 0 & 0 & 1 & 0 & 0 & 0 \\
    0 & 0 & 0 & 0 & 1 & 0 & 0 \\
    0 & 0 & 0 & 0 & 0 & \frac{y}{\sqrt{w^2+y^2}} & 
\frac{w}{\sqrt{w^2+y^2}} \\
    0 & 0 & 0 & 0 & 0 & -\frac{w}{\sqrt{w^2+y^2}} & 
\frac{y}{\sqrt{w^2+y^2}}
      \\
   \end{array}
   \right) \,,
\end{equation}

\noindent we get the HM

\begin{equation}
\label{eh6}
\tilde{h}_7^6=
 \left(
                 \begin{array}{ccccccc}
                  0 & x & 0 & 0 & 0 & 0 & 0 \\
                  x & 0 & z & 0 & 0 & \sqrt{w^2+y^2} & 0 \\
                  0 & z & 0 & z & 0 & 0 & 0 \\
                  0 & 0 & z & 0 & x & \sqrt{w^2+y^2} & 0 \\
                  0 & 0 & 0 & x & 0 & 0 & 0 \\
                  0 & \sqrt{w^2+y^2} & 0 & \sqrt{w^2+y^2} & 0 & 0 & 0 \\
                  0 & 0 & 0 & 0 & 0 & 0 & 0 \\
                 \end{array}
                 \right) \,,
\end{equation}

\noindent with a $6\times 6$ block containing the OHM of the Coutinho graph, noting that $w^2+ y^2 \rightarrow y^2$, see Eq.~\eqref{eq-coutinho-6}. \\

%---------------------------------------

Now, let us assume that the hypothesis holds for $n=k$, {\em i. e.} the OHM of the graph with $6+k$ qubits reduces to the OHM of a chain with only $5$ qubits like the OHM of the Coutinho graph effectively does. The OHM of a graph with $6+(k+1)$ qubits is

\begin{equation}
\label{ehk+1}
h_{k+1}=
\begin{pmatrix}
0&  x&  0 & 0&0&0& \ldots &0  &0 \\ 
x & 0 &  x& 0&0&z_6&z_7 \ldots &  z_{6+k} &  z_{6+k+1} \\
0 &  x&  0 & x &  0 &  0 & \ldots &  0  \\
0 &  0 &  x &0 & x&  z_6&z_7 \ldots  &  z_{6+k} &   z_{6+k+1} \\
0 &  0&0&x&  0 &  0 & \ldots &  0 &0\\
0&z_6&0&z_6 &  0 & 0&\ldots &  0 &0\\
\vdots&\vdots&\vdots&\vdots&\vdots&\vdots&\vdots&\vdots&\vdots\\
0&z_{6+k}&0&z_{6+k} &  0 & 0&\ldots &  0 &0\\
0&z_{6+k+1}&0&z_{6+k+1} &  0 & 0&\ldots &  0 &0
\end{pmatrix} \,,
\end{equation}

\noindent where  we denote with $z_{6+j}$ the links connecting the $6+j$ qubit in the central column of the graph depicted in Figure~\ref{fig-generalizing-coutinho} b) to the second and fourth qubits. 

Defining the GT

\begin{equation}
\label{epk+1}
P_{k+1}=
\begin{pmatrix}
I_{k-1}& & 0& &\\
   &   \frac{z_{6+k}}{\sqrt{z^2_{6+k}+z^2_{6+k+1}}} & & 
 \frac{z_{6+k+1}}{\sqrt{z^2_{6+k}+z^2_{6+k+1}}} \\
0^\dag & & & & \\
    & - \frac{z_{6+k+1}}{\sqrt{z^2_{6+k}+z^2_{6+k+1}}} & & 
\frac{z_{6+k}}{\sqrt{z^2_{6+k}+z^2_{6+k+1}}}  
\end{pmatrix} \,,
\end{equation}

\noindent where $I_{k-1}$ is the $(k-1)\times (k-1)$ identity matrix, $O$ stands for a $(k-1)\times 2$ null matrix,  applying the GT to the OHM in Eq.~\eqref{ehk+1} we get

\begin{equation}
\label{ehk+1gt}
h_{GT}=
\begin{pmatrix}
0&  x&  0 & 0&0&0& \ldots &0  &0 &0\\
x & 0 &  x& 0&0&z_6&z_7 \ldots & z_{6+k-1}& \sqrt{z_{6+k}^2 +  z_{6+k+1}^2} &0\\
0 &  x&  0 & x &  0 &  0 & \ldots &0&0&  0  \\
0 & 0 & x &0 & x& z_6&z_7 \ldots  &z_{6+k-1}& \sqrt{z_{6+k}^2+ z_{6+k+1}^2}&0 \\
0 &  0&0&x&  0 &  0 & \ldots &  0 &0&0\\
0&z_6&0&z_6 &  0 & 0&\ldots &  0 &0&0\\
\vdots&\vdots&\vdots&\vdots&\vdots&\vdots&\vdots&\vdots&\vdots\\
0&z_{6+k-1}&0&z_{6+k-1} &  0 & 0&\ldots &0&  0 &0\\  \\
0&\sqrt{z_{6+k}^2+z_{6+k+1}^2}&0&\sqrt{z_{6+k}^2 +  z_{6+k+1}^2} &  0 & 0&\ldots &0&  0 &0\\
0&0&0&0 &  0 & 0&\ldots &  0 &0&0
\end{pmatrix} \,. 
\end{equation}

\noindent The OHM above contains the $(6+k)\times (6+k)$ block, which is the OHM of a graph with $(6+k)$ qubits and a row and column that are identically null. Note that the strength of one of the coefficients changes after the application of the GT, $z_{6+k} \longrightarrow \sqrt{z_{6+k}^2 + z_{6+k+1} ^2}$. By the inductive hypothesis, choosing the new $\lbrace z_i\rbrace$ adequately, it is possible to obtain PT by continuing the reduction process to get a $5$ qubit chain with PT $\blacksquare$. \\

%---------------------------------

\noindent{\bf Second generalization and Lemma}\\

\begin{figure}[hbt]
\includegraphics[width=0.5\linewidth]{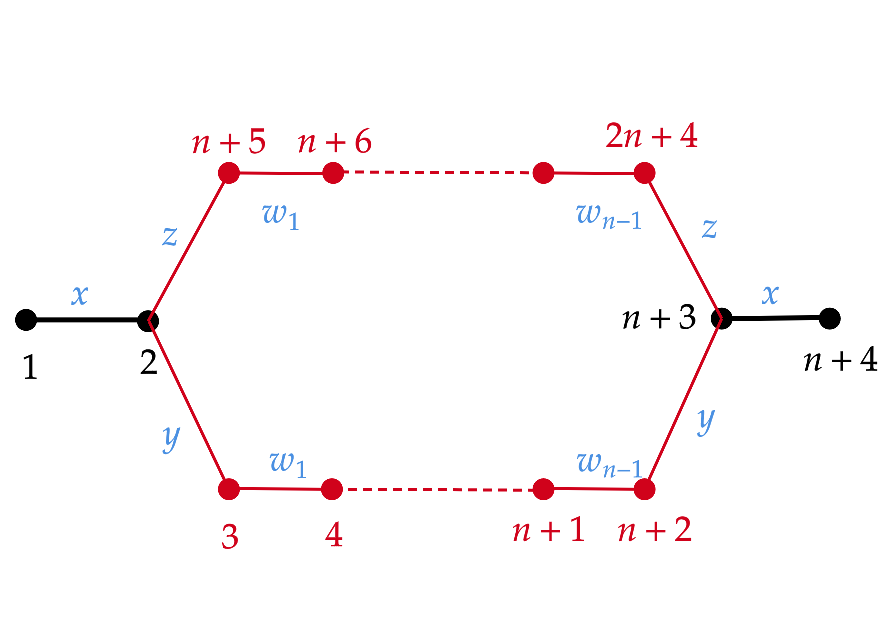}
\caption{The cartoon in the figure shows a generalization of the graph originally proposed by Coutinho, following the prescription given in Fig.~\ref{fig-generalizing-coutinho}(a). The internal set of nodes and links consists of two linear chains with $n$ nodes and $n-1$ links each. The interaction coefficients within this set are labeled from $w_1$ to $w_{n-1}$, while the interaction coefficients corresponding to the links connecting the tails to the chains are labeled $y$ and $z$. The interaction coefficient between the two nodes forming each tail is labeled $x$. See Lemma \ref{lemma2}.\label{fcoutinholarga}. }
\end{figure}

\noindent {\bf Lemma 2. \label{lemma2}} Consider the graph in Figure~\ref{fcoutinholarga}. The graph has $2n+4$ qubits. The OHM of the graph transforms into an OHM that contains a $n+4\times n+4$ block. By choosing the coefficients inside the block adequately,  the OHM of a centrosymmetric qubit chain showing PT is obtained, {\em i. e.} the block is the centrosymmetric tridiagonal matrix for a qubit chain with $n+4$ qubits.

{\bf Proof.}  The graph in Figure~\ref{ftric} b) is equal to the graph in Figure~\ref{fcoutinholarga} for $n=2$. The qubits depicted in red interact with the two neighbours connected by links to them. There are only two qubits that interact with more than two qubits. The OHM is not tridiagonal because of them. The OHM of the graph has two tridiagonal blocks that do not transform under a GT. The matrix is given by

\begin{equation}
\label{emhfcN}
\left(
                 \begin{array}{cccccccccccccccc}
                  0 & x & 0 & 0 & 0 & 0 &\cdots & 0 & 0 & 0 & 0 &\cdots& 0& 0 & 0 & 0 \\
                  x & 0 & y & 0 & 0 & 0 &\cdots& 0 & 0 & 0 & z &\cdots& 0& 0 & 0 & 0 \\
                  0 & y & 0 & w_1 & 0 & 0 &\cdots & 0 & 0 & 0 & 0 &\cdots& 0& 0 & 0 & 0 \\
                  0 & 0 & w_1 & 0 & w_2 & 0 &\cdots & 0 & 0 & 0 & 0 &\cdots& 0& 0 & 0 & 0 \\
                  0 & 0 & 0 & w_2 & 0 & w_3 & \cdots & 0 & 0 & 0 & 0 &\cdots& 0& 0 & 0 & 0 \\
\vdots&\vdots&\vdots&\vdots&\vdots&\vdots&\vdots&\vdots&\vdots&\vdots&\vdots&\vdots&\vdots &\vdots&\vdots&\vdots\\
                  0 & 0 & 0 & 0 &0& 0& \cdots & 0& y & 0 & 0 & \cdots&0 & 0 & 0 & 0 \\
                  0 & 0 & 0 & 0 & 0&0 &\cdots& y & 0 & x & 0 &\cdots& 0 & 0 & 0 & z \\
                  0 & 0 & 0 & 0 & 0 & 0&\cdots& 0 & x & 0 & 0 &\cdots& 0 & 0 & 0 & 0 \\
                  0 & z & 0 & 0 & 0 & 0 &\cdots& 0 & 0 & 0 & 0 &\cdots& 0 & 0&0&0 \\
\vdots&\vdots&\vdots&\vdots&\vdots&\vdots&\vdots&\vdots&\vdots&\vdots&\vdots&\vdots&\vdots &\vdots&\vdots&\vdots\\
                  0 & 0 & 0 & 0 & 0 & 0 &\cdots& 0 & 0 &0&0&\cdots& w_{n-3} & 0 & w_{n-2} & 0 \\
                  0 & 0 & 0 & 0 & 0 & 0 & \cdots&0 & 0&0&0 &\cdots& 0 & w_{n-2} & 0 & w_{n-1} \\
                  0 & 0 & 0 & 0 & 0 & 0 &\cdots & 0 &0& z & 0 &\cdots&0 & 0 & w_{n-1} & 0 \\
                 \end{array}
                 \right) \,,
\end{equation}

\noindent The matrix shows the rows and columns necessary to locate the $z$ coefficients outside the three main diagonals. All the coefficients $x$, $y$, and $\lbrace w_i \rbrace$ lie in the three main diagonals. So, to obtain a tridiagonal matrix, we need to apply a GT that takes the $z$ coefficients into them. Proceeding in the same way that we proceeded with the Coutinho graph, we get the OHM of an  $XX$ chain with $n+4$ qubits whose interaction coefficients are
$J_{12}=x;\;J_{n+3,n+4}=x$, $J_{2,3}=J_{n+2,n+3}=\sqrt{y^2+z^2}$,  and $J_{i,i+1}=w;\;i=3,\ldots,n+1$.\\

\noindent{\bf Third generalization and Theorem}\\

After the considerations about the two graphs generalizing the Coutinho graph, we are in the condition of summarizing both sets of results in a single theorem.

\begin{figure}[hbt]
\includegraphics[width=0.4\linewidth]{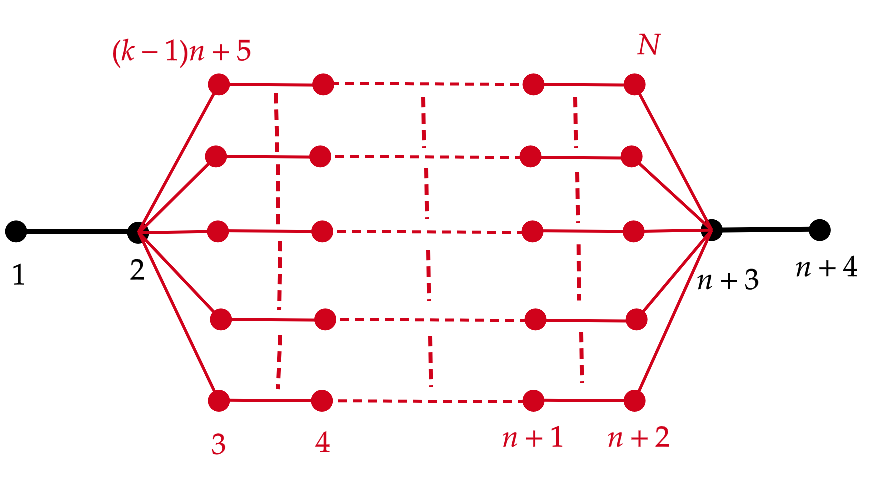}
\caption{The cartoon in the figure shows another generalization in which the set of internal nodes and links consists of $k$ linear chains, each one with $n$ nodes. This graph is the one described in Theorem~\ref{tmr}.\label{fcoutinhog}}
\end{figure}

\noindent {\bf Theorem \label{tmr}}. Consider the qubit graph in Figure~\ref{fcoutinhog}, which contains the two tails formed by the qubit pairs $(1,2)$ and $(n+3,n+4)$. The innermost qubits of both pairs connect with $k$ qubit chains, where each chain has $n$ qubits depicted with red solid dots. This graph reduces to the graphs in Figures~\ref{fcoutinholarga} and \ref{fig-generalizing-coutinho} b) when $k=2$ or when $n=1$, respectively. The OHM of the graph shown in Figure~\ref{fcoutinhog}, which has $N= k n +4$ qubits, reduces to the OHM of an $XX$ chain with $n+4$ qubits, whose parameters, adequately chosen, allow the chain to present PT, in other words, there is always a set of interaction coefficients for the graph such that the graph present PT.

\noindent {\bf Proof.} We can proceed again by mathematical induction. Keeping $n$ fixed, we know that the Theorem holds for $k=2$ by Lemma \ref{lemma2}. So, assuming that it also holds for $k=m$, we need to prove that the Theorem holds for $k=m+1$. The GT needed is similar to the one employed in Lemma \ref{lemma2}, since the OHM of the graph with $k=m+1$ chains has only a pair of interactions extra out of the tridiagonal than the OHM of the graph with $k=m$ chains, let us call them $y_{m+1}$. Using a GT to "eliminate" $y_{m+1}$ renormalizes the strength of the interaction $y_{m}$, and this completes the proof.

\section{Adding extra links to graph that present PT}\label{sec-bridges}

When a graph has loops, like the ones depicted in Figures~\ref{fcou} and \ref{ftric} b), it is instructive to add links between the vertices of the loop. In some cases, the addition modifies the global properties of the graph, as it loses the bipartite condition. In other cases, the additional links allow one to select to which qubits there is perfect quantum transmission or not. We analyse two specific examples in this Section. Besides, note that the graphs in this Section contain paths of different lengths between the qubit where the initial state preparation takes place and the final qubit where the transferred state is retrieved.

\subsection{Wheatstone's bridge}

\begin{figure}[hbt]
\includegraphics[width=0.4\linewidth]{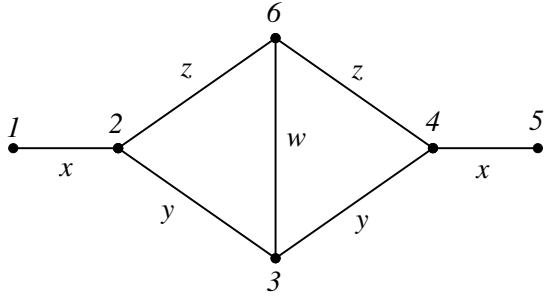}
\caption{The cartoon in the figure shows a graph derived from the one proposed by Coutinho by adding a bridge between nodes 3 and 6. Due to its resemblance to the well-known Wheatstone circuit, we refer to this graph as the Wheatstone bridge. Note that the interaction coefficient corresponding to the bridge is $w$.\label{fpw}}
\end{figure}

We call the graph shown in Figure~\ref{fpw} Wheatstone's bridge because of its likeness to the continuous current circuit that allows measuring the magnitude of electric resistance by balancing the current circulating through the "central wire" connecting nodes $3$ and $6$.

By writing the OHM of the graph using the interaction strengths shown in Figure~\ref{fpw} and eliminating the elements outside the three main diagonals, it is straightforward to show that the OHM obtained has non-zero entries on the main diagonal. In graph theory, the diagonal terms are named "potential terms"  \cite{Kempton2017}, while for magnetic systems, they play the role of an external magnetic field. After performing the necessary GT to obtain a tridiagonal matrix, we found that the six diagonal terms are given by

\[ %\begin{equation}
%\label{epwce}
B_1=B_2=0 \;;\;B_3=\frac{2\,y\,z\,w}{y^2+z^2} \;;\;
B_4=-\frac{2 w^3 y z \left(y^2-z^2\right)^2}{\left(y^2+z^2\right) \left(w^2
    \left(y^2-z^2\right)^2+\left(y^2+z^2\right)^3\right)} \,;
\] %\end{equation}
\[ %\begin{equation}
%\label{epwce}
B_5=
\frac{4 w^3 y z \left(y^2-z^2\right)^2 \left(y^2+z^2\right) \left(w^2
    \left(x^2 \left(y^2-z^2\right)^2-2 y^2 z^2 \left(y^2+z^2\right)\right)+x^2
    \left(y^2+z^2\right)^3\right)}{\left(w^2
    \left(y^2-z^2\right)^2+\left(y^2+z^2\right)^3\right) \left(4 w^4 y^2 z^2
    \left(y^2-z^2\right)^2+w^2 x^2 \left(y^4-z^4\right)^2+x^2
    \left(y^2+z^2\right)^5\right)}\,;
\]
\begin{equation}
\label{epwce}
B_6=
   -\frac{2 w x^2 y z \left(y^2+z^2\right) \left(2 w^2
    \left(y^2-z^2\right)^2+\left(y^2+z^2\right)^3\right)}{4 w^4 y^2 z^2
    \left(y^2-z^2\right)^2+w^2 x^2 \left(y^4-z^4\right)^2+x^2
    \left(y^2+z^2\right)^5}\,.
\end{equation}

\noindent To obtain PT, we search for adequate values for the coefficients $x$, $y$, and $z$ so that the magnetic fields become null. It is clear from the expression of $B_3$ that, except when one of the three variables $x,y$, or $z$ becomes null, there is no way to obtain $B_3=0$. Nevertheless, by choosing $y=z$, we get that the OHM of the graph reads as

\begin{equation}
\label{epwyz}
h\,=\, \left(
                  \begin{array}{cccccc}
                   0 & x & 0 & 0 & 0 & 0 \\
                   x & 0 & \sqrt{2} y & 0 & 0 & 0 \\
                   0 & \sqrt{2} y & w & \sqrt{2} y & 0 & 0 \\
                   0 & 0 & \sqrt{2} y & 0 & x & 0 \\
                   0 & 0 & 0 & x & 0 & 0 \\
                   0 & 0 & 0 & 0 & 0 & -w \\
                  \end{array}
                  \right) \,.
\end{equation}

\noindent Despite the simple structure of the matrix in Eq.~\eqref{epwyz},  its eigenvalues do not satisfy the conditions needed for PT for any non-zero values of $x$, $y$ and $w$.

Let us remember that reducing a given  Hamiltonian matrix to a tridiagonal one allows us to find interaction coefficients compatible with perfect transmission because symmetric tridiagonal matrices are Jacobi matrices. For any spectrum given by $n$ distinct real numbers, there is a Jacobi matrix whose spectrum coincides with that spectrum. In our case, we choose a spectrum compatible with perfect transmission, which determines the Jacobi matrix, which provides the equations for the interaction coefficients since the entries of the Jacobi matrix depend on the interaction coefficients through the GT.

In graph theory, a well-known theorem states that for a homogeneous graph, {\em i.e. }, one where all the interaction coefficients of the $XX$ Hamiltonian are equal to unity with no external field, if $\lambda$ is an eigenvalue with $k$ multiplicity, then $-\lambda$ is also an eigenvalue with $k$ multiplicity if and only if the graph is bipartite \cite{waterloo2022,waterloo2022b}.

We are in a position now to state the main result of the present manuscript. Consider an array of qubits located on the nodes of a bipartite graph with a Hamiltonian with site-dependent interaction coefficients. The graph is centrosymmetric, each node connected to one, two or at most three other nodes. Two nodes with three links can not be directly connected. Then, the OHM of the graph is reducible to the OHM of a linear chain with an $XX$ Hamiltonian, and the graph admits perfect transmission. In what follows, we will continue exploring this statement.

\begin{figure}[hbt]
\includegraphics[width=0.4\linewidth]{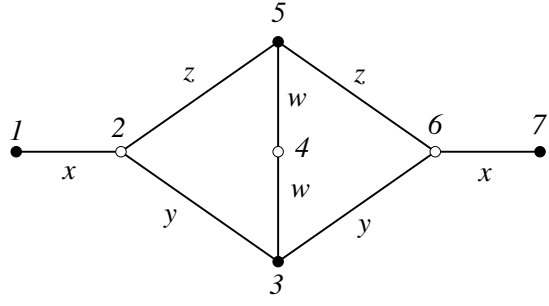}
\caption{The cartoon in the figure shows a generalization of the graph in Fig.~\ref{fpw} obtained by adding a node at the middle of the bridge. This graph is bipartite, in contrast to the one shown in Fig.~\ref{fpw}.
\label{fgpwb}}
\end{figure}

A simple generalisation of Coutinho's graph, including a bridge and preserving the bipartite condition, is the one shown in Figure~\ref{fgpwb}

Applying the GT to the OHM of the graph in Figure~\ref{fgpwb} results in a tridiagonal matrix corresponding to two decoupled linear chains, one with $5$ qubits and the other with only $two$. Imposing again the condition in Eq.~\eqref{eclcspt} to the coupling of the linear chain with $5$ qubits, we get that the coefficients compatible with perfect transmission are given by

\begin{equation}
\label{egpw}
x\,=\,2 \quad ;\quad y\,=\,\sqrt{3} \quad;\quad z\,=\,-y\,=\,-\sqrt{3}
\quad ; \quad w  \, .
\end{equation}

\noindent It is remarkable that this simple example shows that perfect transmission implies that for some graphs, the interaction coefficients can not be all positive. For spin-based qubits, this is equivalent to mixing ferro and antiferro magnetic interactions. The probability of transmission between the first and seventh qubits reads as 

\begin{equation}
\label{epgpwb}
P_{17}(t)\,=\,\sin^4(t) \,.
\end{equation}

\noindent It is also worth pointing out that the probability of transmission is independent of $w$. Nevertheless, imposing the symmetry $z=y$, the existence of PT requires that $w=0$.

\subsection{A key-like graph}

\begin{figure}[hbt]
\includegraphics[width=0.4\linewidth]{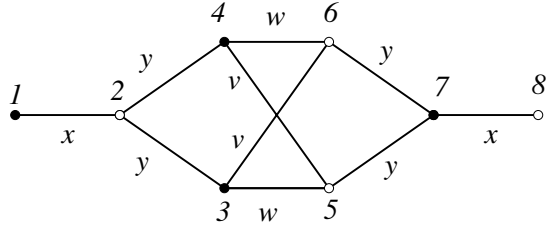}
\caption{The cartoon in the figure shows a graph with maximum distance 5 and two bridges. \label{fcoutinhoN8}}
\end{figure}

There is a modification of the graph depicted in Figure~\ref{ftric} b) that includes the addition of two diagonal bridges. As a consequence, the new graph remains bipartite and has three paths from the initial to the last qubit of the same length. As always, we count the paths without loops. See Figure \ref{fcoutinhoN8}. We start using a coupling set with only four independent coefficients for simplicity.  

After applying appropriate Givens Transformations, the Hamiltonian matrix of the graph in Figure \ref{fcoutinhoN8} has two diagonal blocks, similar to the matrix shown in Eq.~\eqref{emhj4cN8}. These blocks correspond to the Hamiltonian matrices of two decoupled chains with $6$ and $2$ qubits, respectively. 

%The block corresponding to the $6$ qubit chain is given by
%\begin{equation}
%h=h .
%\end{equation}
Using Eq.~\ref{eclcspt} to determine the coupling coefficients, we get that
\begin{equation}
x=\sqrt{5}, \quad \sqrt{2y^2} = \sqrt{8}, \quad \mbox{and} \quad v+w=\sqrt{9} ,
\end{equation}
so the spectrum is given by $\pm 1$, $\pm 3$, and $\pm 5$. The condition $v+w=3$ allows us to find different Hamiltonians with the same probability of transmission from the first qubit to the last one. For $v = w = 3/2$ we get that
\begin{equation}\label{eq-two-bridges}
P_{1,8}(t) = \sin^{10} (t) , \qquad P_{1,1}(t) = \cos^{10} (t),
\end{equation}
where $P_{1,1}(t)$ is the survival probability, {\em i. e.} the probability that the quantum state remains in the first qubit. The probability of transmission in Eq.\ref{eq-two-bridges} is equal to the probability of transmission found for the simple graph without the two bridges, see Eq.\ref{ePcN8}. Interestingly, it is simple to obtain many more exact solutions by asking that the eigenvalues of the block be arbitrary odd integers. In this case, the three positive eigenvalues are given by

\begin{equation}
\lambda_1 = 2j+1 < \lambda_2 = 2k+1 < \lambda_3 = 2m+1, \quad \mbox{with} j, k, m \in \mathbb{N},
\end{equation}

\noindent while the coupling coefficients are given by
\begin{equation}
\label{eccoeffN82b}
x=\sqrt{\frac{(2\,j+1)(2\,k+1)\,(2\,m+1)}{ 2\,(j +  m - k)+1}} \quad;\quad
y=\sqrt{\frac{4\, (j+m+1)\,(k-j)\,(m-k)}{ 2\,(j +  m - k)+1}} \quad;\quad
v\,=w\,=j +  m - k+\frac{1}{2} \,.
\end{equation}

The compatible values for $j=1<k=3<m=5$ result in a probability of transmission that shows PT at an increased frequency,
\begin{equation}
P_{1,8}(t) = \sin^{10}(3t) ,
\end{equation}
which is consistent with the fact that for larger energy eigenvalue values, quantum state transfer appears early on if the eigenvalues satisfy Kay's conditions. 

The construction of different graphs, all showing PT, is allowed by the condition over the values of $j$, $k$, and $m$. For instance, choosing $j=1$, $k=2$, and $m=3$ results in a graph with coupling coefficients given by $x=\sqrt{21}$, $y=2$, and $v=w=5/2$, eigenvalues $3$, $5$, and $7$, and a probability of transmission
\begin{equation}
P_{1,8}(t) = (3+4 \cos(2t))^2 \sin^{10}(t) .
\end{equation}
The coupling coefficients of the equivalent $6$ qubit chain do not satisfy Eq.~\eqref{eclcspt} and are given by
\begin{equation}
J_1= \sqrt{21}, \quad J_2= 2 \sqrt{2}, \quad J_3= 5 .
\end{equation}
\enspace

\noindent {\bf Switching between Perfect Retrieval and Perfect Transmission}

\enspace

The graph shown in Figure~\ref{fcoutinhoN8} changes its transfer properties dramatically if $v=-w$. In this case, irrespective of the values of the coefficients $x$, $y$, and $w$, we get that
\begin{equation}
P_{1,8}(t) = P_{1,5}(t) =P_{1,6}(t)=P_{1,7}(t)=0, \quad \forall t>0 ,
\end{equation}

\noindent that is, the quantum state sweeps between the first four qubits without crossing beyond the qubits at the leftmost extremes of the two bridges. In the absence of decoherence, a quantum state could be stored in this sector of the graph that would work as a memory. At the appropriate moment, changing $v=-3/2$ to $v=3/2$ will send the state to the last qubit. The survival probability when $v=-3/2$ is given by

\begin{equation}
P_{1.1}(v=-3/2,t) = \left(  \frac{8+5 \cos(\sqrt{13} t)}{13} ,
\right)^2 ,
\end{equation}

\noindent which says that the quantum state experiences Perfect Retrieval at times $t_r(n)=2\pi n/\sqrt{13}$. Changing $v=-3/2 \rightarrow 3/2$ at those times produces the perfect transfer from the first qubit to the last one.

\subsection{Chains of superconductor Qubits with control couplings}

\begin{figure}[hbt]
\includegraphics[width=0.5\linewidth]{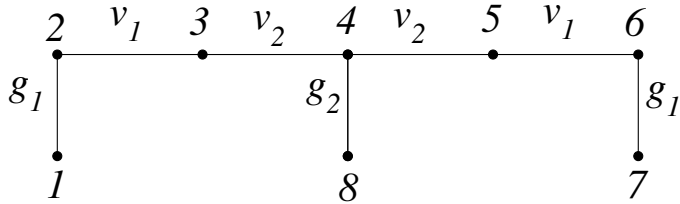}
\caption{In the cartoon of the figure we can observe a decorated linear chain~\cite{Serra2025,Serra2024,Wang2022}.\label{fgtn8}}
\end{figure}

It is interesting to apply the GT procedure to graphs for which it is known that show perfect transmission between some of their sites. For instance, consider the decorated qubit chain depicted in Figure~\ref{fgtn8}. Perfect quantum state transmission is possible between the corner qubits if site-dependent coupling coefficients are allowed \cite{Serra2024,Serra2025}. The corner qubits, in the case shown in Figure~\ref{fgtn8}, are the ones labelled $2$ and $6$. 

Consider the Hamiltonian given by
\begin{equation}
H = \sum_{i=2}^5 \gamma_i (\sigma^+_i \sigma^-_{i+1} + h.c. ) + g_1 (\sigma^+_i1 \sigma^-_{2} + h.c. ) + g_2 (\sigma^+_4 \sigma^-_{8} + h.c. ) + g_3  (\sigma^+_6 \sigma^-_{7} + h.c. )  ,
\end{equation}
where the coefficients $\gamma_i$ are such that $\gamma_2=\gamma_5= v_1$, and \quad $\gamma_3=\gamma_4= v_2$, to match the architecture shown in Fig. \ref{fgtn8}.

The graph is bipartite, and for 
\begin{equation}
g_1 = 1 , \quad g_2=2, \quad v_1 =\sqrt{3}, \quad \mbox{and} \quad v_2= \sqrt{\frac52}  ,
\end{equation}

\noindent We get that the probability of transmission between the corner qubits is given by

\begin{equation}
P_{2,6} = \frac{1}{256} (5 \cos(t) - 8 \cos(2t) + 3 \cos (3 t))^2 .
\end{equation}

\noindent It is simple to show that the spectrum of the OHM 

\begin{equation}
\label{egth}
h\,=\,   \left(
   \begin{array}{cccccccc}
    0 & 1 & 0 & 0 & 0 & 0 & 0 & 0 \\
    1 & 0 & \sqrt{3} & 0 & 0 & 0 & 0 & 0 \\
    0 & \sqrt{3} & 0 & \sqrt{\frac{5}{2}} & 0 & 0 & 0 & 0 \\
    0 & 0 & \sqrt{\frac{5}{2}} & 0 & \sqrt{\frac{5}{2}} & 0 & 0 & 1 \\
    0 & 0 & 0 & \sqrt{\frac{5}{2}} & 0 & \sqrt{3} & 0 & 0 \\
    0 & 0 & 0 & 0 & \sqrt{3} & 0 & 1 & 0 \\
    0 & 0 & 0 & 0 & 0 & 1 & 0 & 0 \\
    0 & 0 & 0 & 1 & 0 & 0 & 0 & 0 \\
   \end{array}
   \right) \,,
\end{equation}

\noindent is given by the eigenvalues $ \lbrace \pm 1, \pm 2, \pm 3\rbrace$, plus a null eigenvalue that is doubly-degenerate. 

After applying the GT to the matrix in Eq.~\eqref{egth}, we get the following Hamiltonian matrix,
\begin{equation}
\label{egth5}
h_5\,=\, G_5 h_4 G_5^t\,=\,
  \left(
   \begin{array}{cccccccc}
    0 & 1 & 0 & 0 & 0 & 0 & 0 & 0 \\
    1 & 0 & \sqrt{3} & 0 & 0 & 0 & 0 & 0 \\
    0 & \sqrt{3} & 0 & \sqrt{\frac{5}{2}} & 0 & 0 & 0 & 0 \\
    0 & 0 & \sqrt{\frac{5}{2}} & 0 & \sqrt{\frac{7}{2}} & 0 & 0 & 0 \\
    0 & 0 & 0 & \sqrt{\frac{7}{2}} & 0 & \sqrt{\frac{15}{7}} & 0 & 0 \\
    0 & 0 & 0 & 0 & \sqrt{\frac{15}{7}} & 0 & \sqrt{\frac{13}{7}} & 0 \\
    0 & 0 & 0 & 0 & 0 & \sqrt{\frac{13}{7}} & 0 & 0 \\
    0 & 0 & 0 & 0 & 0 & 0 & 0 & 0 \\
   \end{array}
   \right)\,.
\end{equation}

\noindent where $h_5$ denotes the OHM obtained after the application of five successive similarity transformations, whose result we denote with $G_5^*$. We present the decomposition of $G_5^*$ in five GTs in Appendix~\ref{a-all-gt-decorated}.

The matrix in Eq.~\ref{egth5} resembles the OHM of a linear $XX$ chain, but it is not centrosymmetric. So, while it is true that asking for the OHM of the equivalent linear $XX$ chain, obtained {\bf after} applying the GT to the OHM of the graph, to fulfil Eq.~\eqref{eclcspt} was instrumental on constructing graphs with PT it is not a necessary condition when the transfer takes place between qubits located "inside" the graph.
Note also that the equivalent linear chain has a length of $7$ qubits since there are only six non-zero coupling coefficients

 \begin{equation}
\label{ejicl}
J_1=1 \quad ; \quad J_2=\sqrt{3} \quad ; \quad J_3=\sqrt{5/2} \quad ; \quad
J_4=\sqrt{7/2} \quad ; \quad
J_5=\sqrt{15/7} \quad ; \quad J_6=\sqrt{13/7} \,.
\end{equation}

Without prior knowledge of the set of coefficients that result in a decorated chain, it is still possible to carry on with the GT procedure for a more general set of coefficients for short-enough decorated chains. For instance, keeping $v_1$ and $v_2$ as variables, and after the corresponding GT, we get for the $8$ qubits decorated chain the HM,

\begin{equation}
\label{egth5g}
h_5\,=\, G_5 h_4 G_5^t\,=\,
   \left(
   \begin{array}{cccccccc}
    0 & 1 & 0 & 0 & 0 & 0 & 0 & 0 \\
    1 & 0 & v_1 & 0 & 0 & 0 & 0 & 0 \\
    0 & v_1 & 0 & v_2 & 0 & 0 & 0 & 0 \\
    0 & 0 & v_2 & 0 & \sqrt{v_2^2+1} & 0 & 0 & 0 \\
    0 & 0 & 0 & \sqrt{v_2^2+1} & 0 & \frac{v_1
      v_2}{\sqrt{v_2^2+1}} & 0 & 0 \\
    0 & 0 & 0 & 0 & \frac{v_1 v_2}{\sqrt{v_2^2+1}} & 0 &
     \sqrt{ \frac{v_1^2+v_2^2+1}{v_2^2+1}} & 0 \\
    0 & 0 & 0 & 0 & 0 &
      \sqrt{ \frac{v_1^2+v_2^2+1}{v_2^2+1}} & 0 & 0 \\
    0 & 0 & 0 & 0 & 0 & 0 & 0 & 0 \\
   \end{array}
   \right) \,.
\end{equation}

\noindent The OHM in Eq.~\eqref{egth5g} is not centrosymmetric but also is equivalent to the OHM of a linear $XX$ chain with seven qubits. 

The matrix in Eq.~\eqref{egth5g} presents succinctly the problems associated with carrying out the GT procedure on the OHM of graphs whose number of qubits with three or more links is on the order of the qubits with only two links. The coupling coefficients of qubits with three or more links would lie outside the $1$-diagonals necessarily. The GTs take a coefficient in, say, the $k$-diagonal and move it until it reaches the $1$-diagonal, modifying some of the coefficients lying in the $1$-diagonal. For instance, taking the matrix coefficient $h\left[4, 8\right]$ in Eq.~\eqref{egth} from the $4$-diagonal up to the $1$-diagonal modifies all the $1$-diagonal coefficients from the fourth up to the eighth row, see Eq.~\eqref{egth5g}. So, if there is no way to avoid the "proliferation" of different coefficients by using symmetry properties, the analytical manipulation becomes very cumbersome. In that case, the matrix coefficients obtained after the GTs consist of quotients of square roots of quadratic polynomials depending on all the different coefficients. For instance, see Eq.~\eqref{egth5g}.

It is instructive to compare the results found for the decorated chain with eight qubits with the results obtained for a longer chain with eleven qubits. For a chain with eleven qubits, there are four $g$-type coefficients, but only two have different values to maintain the symmetry of the chain, $g_1$ and $g_2$, and three $v$-coefficients, $v_1$, $v_2$ and $v_3$. Working out the GTs necessary to reduce the OHM to its tridiagonal form is quite cumbersome if all the coefficients are variables. So, we present results for the following values of the coefficients. Using results found previously \cite{Serra2025}, we know that setting 

\begin{equation}
\label{egtn11}
 g_1\,=\,1  \quad;\quad g_2\,=\,1
\quad ; \quad v_1\,=\,3/\sqrt{2} \quad;\quad v_2\,=\,\sqrt{7/2}
 \quad;\quad v_3\,=\,\sqrt{5} \,,
\end{equation}

\noindent results in the following HM

\begin{equation}
\label{egthN11}
h\,=\,
   \left(
   \begin{array}{ccccccccccc}
    0 & 1 & 0 & 0 & 0 & 0 & 0 & 0 & 0 & 0 & 0 \\
    1 & 0 & \frac{3}{\sqrt{2}} & 0 & 0 & 0 & 0 & 0 & 0 & 0 & 0 \\
    0 & \frac{3}{\sqrt{2}} & 0 & \sqrt{\frac{7}{2}} & 0 & 0 & 0 & 0 & 0 & 0 &
      0 \\
    0 & 0 & \sqrt{\frac{7}{2}} & 0 & \sqrt{5} & 0 & 0 & 0 & 0 & 1 & 0 \\
    0 & 0 & 0 & \sqrt{5} & 0 & \sqrt{5} & 0 & 0 & 0 & 0 & 0 \\
    0 & 0 & 0 & 0 & \sqrt{5} & 0 & \sqrt{\frac{7}{2}} & 0 & 0 & 0 & 1 \\
    0 & 0 & 0 & 0 & 0 & \sqrt{\frac{7}{2}} & 0 & \frac{3}{\sqrt{2}} & 0 & 0 &
      0 \\
    0 & 0 & 0 & 0 & 0 & 0 & \frac{3}{\sqrt{2}} & 0 & 1 & 0 & 0 \\
    0 & 0 & 0 & 0 & 0 & 0 & 0 & 1 & 0 & 0 & 0 \\
    0 & 0 & 0 & 1 & 0 & 0 & 0 & 0 & 0 & 0 & 0 \\
    0 & 0 & 0 & 0 & 0 & 1 & 0 & 0 & 0 & 0 & 0 \\
   \end{array}
   \right) \,  , 
\end{equation}

\noindent which is compatible with perfect transmission. It is simple to show that the spectrum of the OHM in Eq.\eqref{egthN11} is $\lbrace \pm 2 \pm 3, \pm 4, \pm 5\rbrace$, plus a null eigenvalue that is triply-degenerate.

To achieve the tridiagonal form of the HM, it is necessary to employ $14$ different Givens Transformations. Their product is given by
\begin{equation}
\label{egtN11}
G\,=\,\prod_{i=1}^{14}\,G_i \,=\,
   \left(
   \begin{array}{ccccccccccc}
    1 & 0 & 0 & 0 & 0 & 0 & 0 & 0 & 0 & 0 & 0 \\
    0 & 1 & 0 & 0 & 0 & 0 & 0 & 0 & 0 & 0 & 0 \\
    0 & 0 & 1 & 0 & 0 & 0 & 0 & 0 & 0 & 0 & 0 \\
    0 & 0 & 0 & 1 & 0 & 0 & 0 & 0 & 0 & 0 & 0 \\
    0 & 0 & 0 & 0 & \sqrt{\frac{5}{6}} & 0 & 0 & 0 & 0 & \frac{1}{\sqrt{6}} &
      0 \\
    0 & 0 & 0 & 0 & 0 & 1 & 0 & 0 & 0 & 0 & 0 \\
    0 & 0 & 0 & 0 & \frac{\sqrt{\frac{5}{3}}}{8} & 0 &
      \frac{\sqrt{\frac{21}{2}}}{4} & 0 & 0 & -\frac{5}{8 \sqrt{3}} &
      \frac{\sqrt{3}}{4} \\
    0 & 0 & 0 & 0 & 0 & 0 & 0 & 1 & 0 & 0 & 0 \\
    0 & 0 & 0 & 0 & -\frac{3 \sqrt{\frac{35}{163}}}{8} & 0 & \frac{33}{4
      \sqrt{326}} & 0 & \frac{8}{\sqrt{163}} & \frac{15
      \sqrt{\frac{7}{163}}}{8} & -\frac{9 \sqrt{\frac{7}{163}}}{4} \\
    0 & 0 & 0 & 0 & 3 \sqrt{\frac{2}{163}} & 0 &
      -\frac{\sqrt{\frac{35}{163}}}{3} & 0 & \sqrt{\frac{35}{326}} & -3
      \sqrt{\frac{10}{163}} & -\frac{11 \sqrt{\frac{5}{326}}}{3} \\
    0 & 0 & 0 & 0 & 0 & 0 & \frac{1}{3} & 0 & -\frac{1}{\sqrt{2}} & 0 &
      -\frac{\sqrt{\frac{7}{2}}}{3} \\
   \end{array}
  \right)\,.
\end{equation}

Applying the resulting transformation gives the HM.

\begin{equation}
\label{egt14hN11}
h_{14}\,=\, G\,h\,G^t \,=\,
  \left(
   \begin{array}{ccccccccccc}
    0 & 1 & 0 & 0 & 0 & 0 & 0 & 0 & 0 & 0 & 0 \\
    1 & 0 & \frac{3}{\sqrt{2}} & 0 & 0 & 0 & 0 & 0 & 0 & 0 & 0 \\
    0 & \frac{3}{\sqrt{2}} & 0 & \sqrt{\frac{7}{2}} & 0 & 0 & 0 & 0 & 0 & 0 &
      0 \\
    0 & 0 & \sqrt{\frac{7}{2}} & 0 & \sqrt{6} & 0 & 0 & 0 & 0 & 0 & 0 \\
    0 & 0 & 0 & \sqrt{6} & 0 & \frac{5}{\sqrt{6}} & 0 & 0 & 0 & 0 & 0 \\
    0 & 0 & 0 & 0 & \frac{5}{\sqrt{6}} & 0 & \frac{4}{\sqrt{3}} & 0 & 0 & 0 &
      0 \\
    0 & 0 & 0 & 0 & 0 & \frac{4}{\sqrt{3}} & 0 & \frac{3 \sqrt{21}}{8} & 0 & 0
      & 0 \\
    0 & 0 & 0 & 0 & 0 & 0 & \frac{3 \sqrt{21}}{8} & 0 & \frac{\sqrt{163}}{8} &
      0 & 0 \\
    0 & 0 & 0 & 0 & 0 & 0 & 0 & \frac{\sqrt{163}}{8} & 0 & 0 & 0 \\
    0 & 0 & 0 & 0 & 0 & 0 & 0 & 0 & 0 & 0 & 0 \\
    0 & 0 & 0 & 0 & 0 & 0 & 0 & 0 & 0 & 0 & 0 \\
   \end{array}
   \right) \,.
\end{equation}

\section{Discussion and Conclusions}\label{sec-conclusions}

In this work we have shown that Givens transformations provide a simple and systematic way to analyse quantum
state transfer in qubit arrays described by graphs with XX interactions. By applying successive Givens rotations
to the Hamiltonian restricted to the one–excitation subspace, the Hamiltonian matrix of a graph can be reduced
to the tridiagonal form corresponding to an effective linear chain.

This mapping allows us to exploit the well–known results for perfect state transfer in linear XX chains. Once the
graph Hamiltonian is transformed into that of a chain, the conditions for perfect transmission can be imposed on
the chain parameters, which in turn determine the interaction coefficients of the original graph. In this way, the
procedure provides a constructive method to identify families of graphs that exhibit perfect state transfer.

Using this approach we analysed the graph originally introduced by Coutinho and showed how the interaction
coefficients leading to perfect transmission naturally emerge from the mapping to a centrosymmetric linear chain.
We then considered several generalizations of this graph and derived analytic conditions for perfect transmission
in families of graphs containing an arbitrary number of qubits. In particular, we formulated and proved a theorem
that guarantees the existence of interaction coefficients compatible with perfect transmission for graphs formed
by two terminal pairs connected through multiple chains.

We also investigated the effect of adding extra links or bridges to graphs that already present perfect transmission.
These examples illustrate that additional links may destroy the conditions required for perfect transfer or, in some
cases, require the presence of both positive and negative interaction coefficients in order to recover it. Moreover,
our analysis indicates that the bipartite nature of the graph plays an important role in ensuring the existence of
perfect transmission.

Although the method is analytically simple and relies only on elementary matrix transformations, its practical
implementation becomes increasingly cumbersome when the graph contains many vertices with high connectivity
or when the distance between the initial and final qubits becomes large. Nevertheless, the approach provides a
transparent way to relate complex graph architectures with the well understood physics of linear chains.

At a basic level, the requirement that no nodes with three or more links be neighbors of each other arises from the need to keep the nonlinearity of the Givens transformations manageable. In other words, we require that the transformation does not depend on nested square roots, each of which depends quadratically on the coefficients of the Hamiltonian matrix. When the nonlinearity remains simple, finding solutions to the inverse problem is straightforward. Otherwise, solving the resulting set of coupled nonlinear equations for the Hamiltonian coefficients becomes a formidable task.

It is interesting to note that using other methods that reduce the problem to a tridiagonal matrix, such as the Lanczos procedure, one can also obtain a linear chain exhibiting perfect transmission. However, the resulting set of interaction coefficients will not necessarily be the same, since they depend on the specific procedure used. For instance, in the case of the Lanczos method, they depend on the choice of the initial vector for the iterative process. This observation indicates that the inverse problem of determining the interaction coefficients from a given spectrum generally admits multiple possible solutions.

The results presented here open several directions for further work. In particular, it would be interesting to
investigate the robustness of these graph architectures against perturbations of the interaction coefficients and
to explore whether similar constructions can be applied to more general interaction models or to experimental
platforms such as superconducting qubit arrays.

\appendix

\section{GT for the $N=8$ Coutinho graph.}
\label{apn8}

%--------------------------

We devote this Appendix to showing the workings of the Givens transformations over the Hamiltonian Matrix in Eq.~\eqref{emh0cN8}. Looking at the non-zero elements lying below the lower $1$-diagonal, it is easy to identify that there are $z$ coefficients in the second and fifth columns. We begin using the Givens transformation that moves the $(7,2)$ matrix entry toward the lower $1$-diagonal. The Hamiltonian Matrix now reads as

\begin{equation}
\label{emhj2cN8}
   \left(
   \begin{array}{cccccccc}
    0 & x & 0 & 0 & 0 & 0 & 0 & 0 \\
    x & 0 & \sqrt{y^2+z^2}  & 0 & 0 & 0 & 0 & 0 \\
    0 & \sqrt{y^2+z^2}  & 0 & 0 & -\frac{w\,y}{\sqrt{y^2+z^2}} & 0 & 0 & 
\frac{w\,z}{\sqrt{y^2+z^2}}
      \\
    0 & 0 & 0 & 0 & \frac{w\,z}{\sqrt{y^2+z^2}} & 0 & 0 & \frac{w\,y}{\sqrt{y^2+z^2}} \\
    0 & 0 & -\frac{w\,y}{\sqrt{y^2+z^2}} & \frac{w\,z}{\sqrt{y^2+z^2}} & 0 & y & 0 & 0 \\
    0 & 0 & 0 & 0 & y & 0 & x & -z \\
    0 & 0 & 0 & 0 & 0 & x & 0 & 0 \\
    0 & 0 & \frac{w\,z}{\sqrt{y^2+z^2}} & \frac{w\,y}{\sqrt{y^2+z^2}} & 0 & -z & 0 & 0 \\
   \end{array}
   \right)\, .
\end{equation}

\noindent Note that after the transformation, now the OHM has non-zero entries in the third, fourth, fifth, and sixth columns. Also, the entry in the lower $1$-diagonal has changed. 

Applying a GT to the third column gives that the OHM now reads as
\begin{equation}
\label{emhj3cN8}
\left(
                 \begin{array}{cccccccc}
                  0 & x & 0 & 0 & 0 & 0 & 0 & 0 \\
                  x & 0 & \sqrt{y^2+z^2} & 0 & 0 & 0 & 0 & 0 \\
                  0 & \sqrt{y^2+z^2} & 0 & w & 0 & 0 & 0 & 0 \\
                  0 & 0 & w & 0 & 0 & 0 & \sqrt{y^2+z^2} & 0 \\
                  0 & 0 & 0 & 0 & 0 & w & 0 & 0 \\
                  0 & 0 & 0 & 0 & w & 0 & 0 & 0 \\
                  0 & 0 & 0 & \sqrt{y^2+z^2} & 0 & 0 & 0 & x \\
                  0 & 0 & 0 & 0 & 0 & 0 & x & 0 \\
                 \end{array}
                 \right)\,.
\end{equation}

\noindent Note that at this point the only non-zero coefficient below the lower $1$-diagonal is equal to $\sqrt{y^2+z^2}$ and lies on the fourth column. Proceeding as before, we get that
\begin{equation}
\label{emhj4cN8}
 \left(
                 \begin{array}{cccccccc}
                  0 & x & 0 & 0 & 0 & 0 & 0 & 0 \\
                  x & 0 & \sqrt{y^2+z^2} & 0 & 0 & 0 & 0 & 0 \\
                  0 & \sqrt{y^2+z^2} & 0 & w & 0 & 0 & 0 & 0 \\
                  0 & 0 & w & 0 & \sqrt{y^2+z^2} & 0 & 0 & 0 \\
                  0 & 0 & 0 & \sqrt{y^2+z^2} & 0 & 0 & 0 & x \\
                  0 & 0 & 0 & 0 & 0 & 0 & w & 0 \\
                  0 & 0 & 0 & 0 & 0 & w & 0 & 0 \\
                  0 & 0 & 0 & 0 & x & 0 & 0 & 0 \\
                 \end{array}
                 \right)\,.
\end{equation}

\noindent In this iteration of the HM, the non-zero element outside the $1$ diagonals is an $x$ in the fifth column. Applying the corresponding GT, we finally get the tridiagonal matrix

\begin{equation}
\label{emhfcN8}
\left(
\begin{array}{cccccccc}
                  0 & x & 0 & 0 & 0 & 0 & 0 & 0 \\
                  x & 0 & \sqrt{y^2+z^2} & 0 & 0 & 0 & 0 & 0 \\
                  0 & \sqrt{y^2+z^2} & 0 & w & 0 & 0 & 0 & 0 \\
                  0 & 0 & w & 0 & \sqrt{y^2+z^2} & 0 & 0 & 0 \\
                  0 & 0 & 0 & \sqrt{y^2+z^2} & 0 & x & 0 & 0 \\
                  0 & 0 & 0 & 0 & x & 0 & 0 & 0 \\
                  0 & 0 & 0 & 0 & 0 & 0 & 0 & w \\
                  0 & 0 & 0 & 0 & 0 & 0 & w & 0 \\
                 \end{array}
                 \right)\,,
\end{equation}

%-------------------------------

\section{The Given transformations employed on the Hamiltonian Matrix of the decorated chain with eight qubits}
\label{a-all-gt-decorated}
%-----------------------------------------

Starting from the matrix in Eq.~\eqref{egth}, the first GT, $G_1$, is designed to move $h_{4,8}$ closer to the corresponding $1$-diagonal. The matrix implementing $G_1$ is given by
\begin{equation}
\label{egt1}
G_1\,=\,
 \left(
                 \begin{array}{cccccccc}
                  1 & 0 & 0 & 0 & 0 & 0 & 0 & 0 \\
                  0 & 1 & 0 & 0 & 0 & 0 & 0 & 0 \\
                  0 & 0 & 1 & 0 & 0 & 0 & 0 & 0 \\
                  0 & 0 & 0 & 1 & 0 & 0 & 0 & 0 \\
                  0 & 0 & 0 & 0 & 1 & 0 & 0 & 0 \\
                  0 & 0 & 0 & 0 & 0 & 1 & 0 & 0 \\
                  0 & 0 & 0 & 0 & 0 & 0 & 0 & 1 \\
                  0 & 0 & 0 & 0 & 0 & 0 & -1 & 0 \\
                 \end{array}
                 \right) \, .
\end{equation}

\noindent We see the effect of the similarity transformation on the Hamiltonian in the following Equation

\begin{equation}
\label{egth1}
h_1\,=\, G_1 h G_1^t\,=\,
   \left(
   \begin{array}{cccccccc}
    0 & 1 & 0 & 0 & 0 & 0 & 0 & 0 \\
    1 & 0 & \sqrt{3} & 0 & 0 & 0 & 0 & 0 \\
    0 & \sqrt{3} & 0 & \sqrt{\frac{5}{2}} & 0 & 0 & 0 & 0 \\
    0 & 0 & \sqrt{\frac{5}{2}} & 0 & \sqrt{\frac{5}{2}} & 0 & 1 & 0 \\
    0 & 0 & 0 & \sqrt{\frac{5}{2}} & 0 & \sqrt{3} & 0 & 0 \\
    0 & 0 & 0 & 0 & \sqrt{3} & 0 & 0 & -1 \\
    0 & 0 & 0 & 1 & 0 & 0 & 0 & 0 \\
    0 & 0 & 0 & 0 & 0 & -1 & 0 & 0 \\
   \end{array}
   \right) \,.
\end{equation}

\noindent Note that now $(h_1)_{4,8}=0$, but $(h_1)_{4,7}=1$. The Givens Transformation that takes $(h_1)_{4,7} \longrightarrow 0$ reads as
\begin{equation}
\label{egt2}
G_2\,=\,
\left(
                 \begin{array}{cccccccc}
                  1 & 0 & 0 & 0 & 0 & 0 & 0 & 0 \\
                  0 & 1 & 0 & 0 & 0 & 0 & 0 & 0 \\
                  0 & 0 & 1 & 0 & 0 & 0 & 0 & 0 \\
                  0 & 0 & 0 & 1 & 0 & 0 & 0 & 0 \\
                  0 & 0 & 0 & 0 & 1 & 0 & 0 & 0 \\
                  0 & 0 & 0 & 0 & 0 & 0 & 1 & 0 \\
                  0 & 0 & 0 & 0 & 0 & -1 & 0 & 0 \\
                  0 & 0 & 0 & 0 & 0 & 0 & 0 & 1 \\
                 \end{array}
                 \right) \, . 
\end{equation}

\noindent Now the similarity transformation produces a new OHM with the same spectrum that reads as
\begin{equation}
\label{egth2}
h_2\,=\, G_2 h_1 G_2^t\,=\,
  \left(
   \begin{array}{cccccccc}
    0 & 1 & 0 & 0 & 0 & 0 & 0 & 0 \\
    1 & 0 & \sqrt{3} & 0 & 0 & 0 & 0 & 0 \\
    0 & \sqrt{3} & 0 & \sqrt{\frac{5}{2}} & 0 & 0 & 0 & 0 \\
    0 & 0 & \sqrt{\frac{5}{2}} & 0 & \sqrt{\frac{5}{2}} & 1 & 0 & 0 \\
    0 & 0 & 0 & \sqrt{\frac{5}{2}} & 0 & 0 & -\sqrt{3} & 0 \\
    0 & 0 & 0 & 1 & 0 & 0 & 0 & 0 \\
    0 & 0 & 0 & 0 & -\sqrt{3} & 0 & 0 & 1 \\
    0 & 0 & 0 & 0 & 0 & 0 & 1 & 0 \\
   \end{array}
   \right) \, .
\end{equation}

The other GT needed are $G_3$, $G_4$, and $G_5$, which are given by
\begin{equation}
\label{egt3}
G_3\,=\,
  \left(
   \begin{array}{cccccccc}
    1 & 0 & 0 & 0 & 0 & 0 & 0 & 0 \\
    0 & 1 & 0 & 0 & 0 & 0 & 0 & 0 \\
    0 & 0 & 1 & 0 & 0 & 0 & 0 & 0 \\
    0 & 0 & 0 & 1 & 0 & 0 & 0 & 0 \\
    0 & 0 & 0 & 0 & \sqrt{\frac{5}{7}} & \sqrt{\frac{2}{7}} & 0 & 0 \\
    0 & 0 & 0 & 0 & -\sqrt{\frac{2}{7}} & \sqrt{\frac{5}{7}} & 0 & 0 \\
    0 & 0 & 0 & 0 & 0 & 0 & 1 & 0 \\
    0 & 0 & 0 & 0 & 0 & 0 & 0 & 1 \\
   \end{array}
   \right) \,,
\end{equation}

\begin{equation}
\label{egt4}
G_4\,=\,
\left(
                  \begin{array}{cccccccc}
                   1 & 0 & 0 & 0 & 0 & 0 & 0 & 0 \\
                   0 & 1 & 0 & 0 & 0 & 0 & 0 & 0 \\
                   0 & 0 & 1 & 0 & 0 & 0 & 0 & 0 \\
                   0 & 0 & 0 & 1 & 0 & 0 & 0 & 0 \\
                   0 & 0 & 0 & 0 & 1 & 0 & 0 & 0 \\
                   0 & 0 & 0 & 0 & 0 & 0 & -1 & 0 \\
                   0 & 0 & 0 & 0 & 0 & 1 & 0 & 0 \\
                   0 & 0 & 0 & 0 & 0 & 0 & 0 & 1 \\
                  \end{array}
                  \right) \,,
\end{equation}

\noindent and 
\begin{equation}
\label{egt5}
G_5\,=\,
  \left(
   \begin{array}{cccccccc}
    1 & 0 & 0 & 0 & 0 & 0 & 0 & 0 \\
    0 & 1 & 0 & 0 & 0 & 0 & 0 & 0 \\
    0 & 0 & 1 & 0 & 0 & 0 & 0 & 0 \\
    0 & 0 & 0 & 1 & 0 & 0 & 0 & 0 \\
    0 & 0 & 0 & 0 & 1 & 0 & 0 & 0 \\
    0 & 0 & 0 & 0 & 0 & 1 & 0 & 0 \\
    0 & 0 & 0 & 0 & 0 & 0 & -\sqrt{\frac{6}{13}} & -\sqrt{\frac{7}{13}} \\
    0 & 0 & 0 & 0 & 0 & 0 & \sqrt{\frac{7}{13}} & -\sqrt{\frac{6}{13}} \\
   \end{array}
   \right) \, .
\end{equation}

\noindent The OHM obtained after applying all the GTs, $G_1, \ldots, G_5$ is shown in Eq.~\ref{egth5}.

%---------------------------------

%\bibliographystyle{plainurl}
%\bibliographystyle{unsrtdin}
\bibliography{trans-givens}

@article{Xie2023,
  title = {Breaking the speed limit for perfect quantum state transfer},
  author = {Xie, Weichen and Kay, Alastair and Tamon, Christino},
  journal = {Phys. Rev. A},
  volume = {108},
  issue = {1},
  pages = {012408},
  numpages = {7},
  year = {2023},
  month = {Jul},
  publisher = {American Physical Society},
  doi = {10.1103/PhysRevA.108.012408},
  url = {https://link.aps.org/doi/10.1103/PhysRevA.108.012408}
}

@article{Maleki2021,
title = {Perfect swap and transfer of arbitrary quantum states},
journal = {Optics Communications},
volume = {496},
pages = {126870},
year = {2021},
issn = {0030-4018},
doi = {https://doi.org/10.1016/j.optcom.2021.126870},
url = {https://www.sciencedirect.com/science/article/pii/S0030401821001206},
author = {Yusef Maleki and Aleksei M. Zheltikov}
}

@article{Mograby2021,
doi = {10.1088/1751-8121/abc4b9},
url = {https://dx.doi.org/10.1088/1751-8121/abc4b9},
year = {2021},
month = {mar},
publisher = {IOP Publishing},
volume = {54},
number = {12},
pages = {125301},
author = {Mograby, Gamal and Derevyagin, Maxim and Dunne, Gerald V and Teplyaev, Alexander},
title = {Spectra of perfect state transfer Hamiltonians on fractal-like graphs},
journal = {Journal of Physics A: Mathematical and Theoretical}
}

@article{Christandl2004,
  title = {Perfect State Transfer in Quantum Spin Networks},
  author = {Christandl, Matthias and Datta, Nilanjana and Ekert, Artur and Landahl, Andrew J.},
  journal = {Phys. Rev. Lett.},
  volume = {92},
  issue = {18},
  pages = {187902},
  numpages = {4},
  year = {2004},
  month = {May},
  publisher = {American Physical Society},
  doi = {10.1103/PhysRevLett.92.187902},
  url = {https://link.aps.org/doi/10.1103/PhysRevLett.92.187902}
}

@article{Li2018,
title = {Perfect Quantum State Transfer in a Superconducting Qubit Chain with Parametrically Tunable Couplings},
 author = {Li, X. and Ma, Y. and Han, J. and Chen, Tao and Xu, Y. and Cai, W. and Wang, H. and Song, Y.P. and Xue, Zheng-Yuan and Yin, Zhang-qi and Sun, Luyan},
 journal = {Phys. Rev. Appl.},
 volume = {10},
 issue = {5},
 pages = {054009},
 numpages = {11},
 year = {2018},
 month = {Nov},
 publisher = {American Physical Society},
 doi = {10.1103/PhysRevApplied.10.054009},
 url = {https://link.aps.org/doi/10.1103/PhysRevApplied.10.054009}
}

@article{Bayat2014,
 title = {Arbitrary perfect state transfer in $d$-level spin chains},
 author = {Bayat, Abolfazl},
 journal = {Phys. Rev. A},
 volume = {89},
 issue = {6},
 pages = {062302},
 numpages = {6},
 year = {2014},
 month = {Jun},
 publisher = {American Physical Society},
 doi = {10.1103/PhysRevA.89.062302},
 url = {https://link.aps.org/doi/10.1103/PhysRevA.89.062302}
}

@article{Godsil_2011,
  author       = {Chris D. Godsil},
  title        = {Periodic Graphs},
  journal      = {The Electronic Journal of Combinatorics},
  volume       = {18},
  number       = {1},
  pages        = {P23},
  year         = {2011},
  url          = {https://doi.org/10.37236/510},
  doi          = {https://doi.org/10.37236/510},
}

@article{Fan2013,
title = {Pretty good state transfer on double stars},
journal = {Linear Algebra and its Applications},
volume = {438},
number = {5},
pages = {2346-2358},
year = {2013},
issn = {0024-3795},
doi = {https://doi.org/10.1016/j.laa.2012.10.006},
url = {https://www.sciencedirect.com/science/article/pii/S0024379512007161},
author = {Xiaoxia Fan and Chris Godsil}
}

@article{Kempton2017,
author = {Kempton, Mark and Lippner, Gabor and Yau, Shing-Tung},
title = {Perfect state transfer on graphs with a potential},
year = {2017},
issue_date = {March 2017},
publisher = {Rinton Press, Incorporated},
address = {Paramus, NJ},
volume = {17},
number = {3–4},
issn = {1533-7146},
journal = {Quantum Info. Comput.},
month = mar,
pages = {303–327},
numpages = {25}
}

@article{Cheung2011,
title = {Perfect state transfer in cubelike graphs},
journal = {Linear Algebra and its Applications},
volume = {435},
number = {10},
pages = {2468-2474},
year = {2011},
note = {Special Issue in Honor of Dragos Cvetkovic},
issn = {0024-3795},
doi = {https://doi.org/10.1016/j.laa.2011.04.022},
url = {https://www.sciencedirect.com/science/article/pii/S0024379511003272},
author = {Wang-Chi Cheung and Chris Godsil}
}

@misc{kay2018,
      title={The Perfect State Transfer Graph Limbo}, 
      author={Alastair Kay},
      year={2018},
      eprint={1808.00696},
      archivePrefix={arXiv},
      primaryClass={quant-ph},
      url={https://arxiv.org/abs/1808.00696}, 
}

@article{kay2010,
author = {Kay, Alastair},
title = {Perfect, efficient, state transfer and its application as a cconstructive tool},
journal = {International Journal of Quantum Information},
volume = {08},
number = {04},
pages = {641-676},
year = {2010},
doi = {10.1142/S0219749910006514},
URL = { https://doi.org/10.1142/S0219749910006514},
eprint = {https://doi.org/10.1142/S0219749910006514}
}

@article{Ge2011,
author = {GE, YANG and GREENBERG, BENJAMIN and PEREZ, OSCAR and TAMON, CHRISTINO},
title = {PERFECT STATE TRANSFER, GRAPH PRODUCTS AND EQUITABLE PARTITIONS},
journal = {International Journal of Quantum Information},
volume = {09},
number = {03},
pages = {823-842},
year = {2011},
doi = {10.1142/S0219749911007472},
URL = {https://doi.org/10.1142/S0219749911007472},
eprint = {https://doi.org/10.1142/S0219749911007472}
}

@article{acde2004,
  title = {Mirror Inversion of Quantum States in Linear Registers},
  author = {Albanese, Claudio and Christandl, Matthias and Datta, Nilanjana and Ekert, Artur},
  journal = {Phys. Rev. Lett.},
  volume = {93},
  issue = {23},
  pages = {230502},
  numpages = {4},
  year = {2004},
  month = {Nov},
  publisher = {American Physical Society},
  doi = {10.1103/PhysRevLett.93.230502},
  url = {https://link.aps.org/doi/10.1103/PhysRevLett.93.230502}
}

@Book{numericalrecipes,
  author    = "William H. Press and Brian P. Flannery and  Saul A. Teukolsky and William T. Vetterling",
  title     = "Numerical Recipes in FORTRAN 77",
  publisher = "Cambridge University Press",
  year      = "1992",
  volume    = "1",
  address   = "Cambridge, UK",
  edition   = "1",
  month     = "September",
  note      = ""
}

@BOOK{lapack,
      AUTHOR = {Anderson, E. and Bai, Z. and Bischof, C. and
                Blackford, S. and Demmel, J. and Dongarra, J. and
                Du Croz, J. and Greenbaum, A. and Hammarling, S. and
                McKenney, A. and Sorensen, D.},
      TITLE = {{LAPACK} Users' Guide},
      EDITION = {Third},
      PUBLISHER = {Society for Industrial and Applied Mathematics},
      YEAR = {1999},
      ADDRESS = {Philadelphia, PA},
      ISBN = {0-89871-447-8 (paperback)} }

@article{coutinho2016,
title = {Tridiagonalization of a symmetric matrix on a square array of mesh-connected processors},
journal = {The Electronic Journal of Combinatorics},
volume = {23},
number = {1},
pages = {46},
year = {2016},
doi = {https://doi.org/10.37236/5031 },
url = {https://www.combinatorics.org/ojs/index.php/eljc/article/view/v23i1p46},
author = {Gabriel Coutinho},
}

@book{waterloo2022,
  author = {Andries E. Brouwer and  Willem H. Haemers},
  year = {2012},
  title = {Spectra of Graphs},
  publisher = {Springer},
  address = {New York, NY},
}

@misc{waterloo2022b,
  author = {University of Waterloo},
  title = {CS 860: Eigenvalues and Polynomials},
  url = {https://cs.uwaterloo.ca/~lapchi/cs860-2022/notes/eigenpoly.pdf},
  year = {2022},
}

@article{Serra2025,
title = {Perfect, Pretty Good and optimised Quantum State Transfer in Transmon qubit
chains},
journal = {Physics Letters A},
volume = {25},
pages = {239-253},
year = {2025},
issn = {0024-3795},
doi = {https://doi.org/10.1016/0024-3795(79)90021-1},
url = {https://www.sciencedirect.com/science/article/pii/0024379579900211},
author = {Serra, Pablo and Ferrón, Alejandro and Osenda, Omar}
}

@article{Bose2003,
  title = {Quantum Communication through an Unmodulated Spin Chain},
  author = {Bose, Sougato},
  journal = {Phys. Rev. Lett.},
  volume = {91},
  issue = {20},
  pages = {207901},
  numpages = {4},
  year = {2003},
  month = {Nov},
  publisher = {American Physical Society},
  doi = {10.1103/PhysRevLett.91.207901},
  url = {https://link.aps.org/doi/10.1103/PhysRevLett.91.207901}
}

@article{Bose2008,
author = {Sougato Bose},
title = {Quantum communication through spin chain dynamics: an introductory overview},
journal = {Contemporary Physics},
volume = {48},
number = {1},
pages = {13--30},
year = {2007},
publisher = {Taylor \& Francis},
doi = {10.1080/00107510701342313},
URL = {https://doi.org/10.1080/00107510701342313},
eprint = {https://doi.org/10.1080/00107510701342313}
}

@article{Serra2022JPA,
doi = {10.1088/1751-8121/ac901d},
url = {https://dx.doi.org/10.1088/1751-8121/ac901d},
year = {2022},
month = {sep},
publisher = {IOP Publishing},
volume = {55},
number = {40},
pages = {405302},
author = {Serra, Pablo and Ferrón, Alejandro and Osenda, Omar},
title = {Exact solution of a family of staggered Heisenberg chains with conclusive pretty good quantum state transfer},
journal = {Journal of Physics A: Mathematical and Theoretical}
}

@article{Wang2022,
  title = {Arbitrary entangled state transfer via a topological qubit chain},
  author = {Wang, Chong and Li, Linhu and Gong, Jiangbin and Liu, Yu-xi},
  journal = {Phys. Rev. A},
  volume = {106},
  issue = {5},
  pages = {052411},
  numpages = {19},
  year = {2022},
  month = {Nov},
  publisher = {American Physical Society},
  doi = {10.1103/PhysRevA.106.052411},
  url = {https://link.aps.org/doi/10.1103/PhysRevA.106.052411}
}

@article{Yousefjani2021a,
  doi = {10.22331/q-2021-05-26-460},
  url = {https://doi.org/10.22331/q-2021-05-26-460},
  title = {Parallel entangling gate operations and two-way quantum communication in spin chains},
  author = {Yousefjani, Rozhin and Bayat, Abolfazl},
  journal = {{Quantum}},
  issn = {2521-327X},
  publisher = {{Verein zur F{\"{o}}rderung des Open Access Publizierens in den Quantenwissenschaften}},
  volume = {5},
  pages = {460},
  month = may,
  year = {2021}
}

@article{Zwick2015,
  title        = {Quantum state transfer in disordered spin chains: How much engineering is reasonable?},
  author       = {Zwick, Analia and Alvarez, Gonzalo A. and Stolze, Joachim and Osenda,Omar},
  year         = {2015},
  month        = {May},
  journal      = {Quantum Information \& Computation},
  volume       = {15},
  number       = {7\& 8},
  pages        = {0582--0600},
  publisher    = {Rington Press},
  url="https://www.rintonpress.com/xxqic15/qic-15-78/0582-0600.pdf"
}

@article{Serra2024,
doi = {10.1088/1751-8121/ad0d20},
url = {https://dx.doi.org/10.1088/1751-8121/ad0d20},
year = {2023},
month = {dec},
publisher = {IOP Publishing},
volume = {57},
number = {1},
pages = {015304},
author = {Serra, Pablo and Ferrón, Alejandro and Osenda, Omar},
title = {The scaling law of the arrival time of spin systems that present pretty good transmission},
journal = {Journal of Physics A: Mathematical and Theoretical}
}

@article{Bravyi2022,
    author = {Bravyi, Sergey and Dial, Oliver and Gambetta, Jay M. and Gil, Darío and Nazario, Zaira},
    title = {The future of quantum computing with superconducting qubits},
    journal = {Journal of Applied Physics},
    volume = {132},
    number = {16},
    pages = {160902},
    year = {2022},
    month = {10},
    issn = {0021-8979},
    doi = {10.1063/5.0082975},
    url = {https://doi.org/10.1063/5.0082975},
    eprint = {https://pubs.aip.org/aip/jap/article-pdf/doi/10.1063/5.0082975/20034201/160902\_1\_5.0082975.pdf},
}

@misc{heavyhex,
  author = {Nation, Paul and Paik, Hanhee and Cross, Andrew and Nazario, Zaira},
  title = {The IBM Quantum Heavy Hex Lattice},
  url = {https://research.ibm.com/blog/heavy-hex-lattice},
  year = {2022},
}

@article{Kattemolle2025,
author = {Kattem\"{o}lle, Joris and Hariharan, Seenivasan},
title = {Line-Graph Qubit Routing},
year = {2025},
issue_date = {September 2025},
publisher = {Association for Computing Machinery},
address = {New York, NY, USA},
volume = {6},
number = {3},
url = {https://doi.org/10.1145/3733842},
doi = {10.1145/3733842},
journal = {ACM Transactions on Quantum Computing},
month = jun,
articleno = {22},
numpages = {18}
}

@article{Karbach2005,
  title = {Spin chains as perfect quantum state mirrors},
  author = {Karbach, Peter and Stolze, Joachim},
  journal = {Phys. Rev. A},
  volume = {72},
  issue = {3},
  pages = {030301},
  numpages = {4},
  year = {2005},
  month = {Sep},
  publisher = {American Physical Society},
  doi = {10.1103/PhysRevA.72.030301},
  url = {https://link.aps.org/doi/10.1103/PhysRevA.72.030301}
}

@incollection{VENKATESHAN201419,
title = {Chapter 2 - Solution of Linear Equations},
editor = {S.P. Venkateshan and Prasanna Swaminathan},
booktitle = {Computational Methods in Engineering},
publisher = {Academic Press},
address = {Boston},
pages = {19-103},
year = {2014},
isbn = {978-0-12-416702-5},
doi = {https://doi.org/10.1016/B978-0-12-416702-5.50002-8},
url = {https://www.sciencedirect.com/science/article/pii/B9780124167025500028},
author = {S.P. Venkateshan and Prasanna Swaminathan}
}

@article{Coutinho2022,
title = {Quantum walks do not like bridges},
journal = {Linear Algebra and its Applications},
volume = {652},
pages = {155-172},
year = {2022},
issn = {0024-3795},
doi = {https://doi.org/10.1016/j.laa.2022.07.009},
url = {https://www.sciencedirect.com/science/article/pii/S0024379522002671},
author = {Gabriel Coutinho and Chris Godsil and Emanuel Juliano and Christopher M. {van Bommel}}
}

@article{Coutinho2015,
author = {Coutinho, Gabriel and Liu, Henry},
title = {No Laplacian Perfect State Transfer in Trees},
journal = {SIAM Journal on Discrete Mathematics},
volume = {29},
number = {4},
pages = {2179-2188},
year = {2015},
doi = {10.1137/140989510},
URL = { https://doi.org/10.1137/140989510},
eprint = {https://doi.org/10.1137/140989510}
}

@misc{Himmel2025,
      title={State Transfer in Latent-Symmetric Networks}, 
      author={Jonas Himmel and Max Ehrhardt and Matthias Heinrich and Sebastian Weidemann and Tom A. W. Wolterink and Malte Röntgen and Peter Schmelcher and Alexander Szameit},
      year={2025},
      eprint={2501.12185},
      archivePrefix={arXiv},
      primaryClass={quant-ph},
      url={https://arxiv.org/abs/2501.12185}, 
}

@misc{Himmel2025b,
      title={Eigenmodes of latent-symmetric quantum photonic networks}, 
      author={Jonas Himmel and Max Ehrhardt and Matthias Heinrich and Malte Röntgen and Alexander Szameit and Tom A. W. Wolterink},
      year={2025},
      eprint={2501.13029},
      archivePrefix={arXiv},
      primaryClass={physics.optics},
      url={https://arxiv.org/abs/2501.13029}, 
}

@inproceedings{Ma2024,
author = {Ma, Xinyu and Chu, Xu and Yang, Zhibang and Lin, Yang and Gao, Xin and Zhao, Junfeng},
title = {Parameter efficient quasi-orthogonal fine-tuning via givens rotation},
year = {2024},
publisher = {JMLR.org},
booktitle = {Proceedings of the 41st International Conference on Machine Learning},
articleno = {1369},
numpages = {44},
location = {Vienna, Austria},
series = {ICML'24}
}

@ARTICLE{Cybenko2001,
author={Cybenko, George},
journal={ Computing in Science \& Engineering },
title={{ Reducing Quantum Computations to Elementary Unitary Operations }},
year={2001},
volume={3},
number={02},
ISSN={1558-366X},
pages={27-32},
doi={10.1109/5992.908999},
url = {https://doi.ieeecomputersociety.org/10.1109/5992.908999},
publisher={IEEE Computer Society},
address={Los Alamitos, CA, USA},
month=mar}

\end{document}